\documentclass[conference]{IEEEtran}

\usepackage{tikz}
\usepackage{amsmath}
\usepackage{booktabs}
\usepackage{multirow}
\usepackage{lipsum}
\usepackage{multicol}
\usepackage{cleveref}
\usepackage{rotating}
\usepackage{pifont}
\crefname{section}{§}{§§}
\Crefname{section}{§}{§§}

\ifCLASSINFOpdf
\else
\fi

\begin{document}


\title{Investigating co-occurrences of MITRE ATT\&CK Techniques}

\author{Md Rayhanur Rahman, Laurie Williams}


\maketitle

\begin{abstract}
Cyberattacks use adversarial techniques to bypass system defenses, persist, and eventually breach systems. The MITRE ATT\&CK framework catalogs a set of adversarial techniques and maps between adversaries and their used techniques and tactics. Understanding how adversaries deploy techniques in conjunction is pivotal for learning adversary behavior, hunting potential threats, and formulating a proactive defense. \textit{The goal of this research is to aid cybersecurity practitioners and researchers in choosing detection and mitigation strategies through co-occurrence analysis of adversarial techniques reported in MITRE ATT\&CK.} We collect the adversarial techniques of 115 cybercrime groups and 484 malware from the MITRE ATT\&CK. We apply association rule mining and network analysis to investigate how adversarial techniques co-occur. We identify that adversaries pair \textit{T1059: Command and scripting interface} and \textit{T1105: Ingress tool transfer} techniques with a relatively large number of ATT\&CK techniques. We also identify adversaries using the \textit{T1082: System Information Discovery} technique to determine their next course of action. We observe adversaries deploy the highest number of techniques from the \textit{TA0005: Defense evasion} and \textit{TA0007: Discovery} tactics. Based on our findings on co-occurrence, we identify six detection, six mitigation strategies, and twelve adversary behaviors. We urge defenders to prioritize primarily the detection of \textit{TA0007: Discovery} and mitigation of \textit{TA0005: Defense evasion} techniques. Overall, this study approximates how adversaries leverage techniques based on publicly reported documents. We advocate organizations investigate adversarial techniques in their environment and make the findings available for a more precise and actionable understanding. 
\end{abstract}

\section{Introduction}

\begin{figure}
    \centering
    \includegraphics[width=\columnwidth]{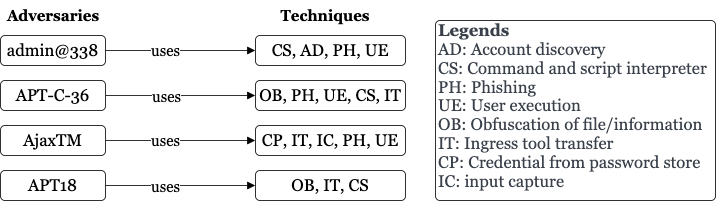}
    \caption{Example of four real-world cyberattacks consisting of multiple techniques~\cite{admin338, aptC36, ajaxtm, apt18}}
    \label{fig:attack-example}
\end{figure}

Information technology (IT) systems draw continuous attention from threat actors with financial motives~\cite{hackernews-financial-backup} and organized backing~\cite{gaoHinCTICyberThreat2022}. In 2021, corporate businesses suffered 50\% more cyberattacks per week compared to 2020~\cite{checkpoint-research}. The incurred cost of cyberattacks also keeps rising. Cybersecurity Ventures forecasts that financial damage by cyberattack will be \$6,000B in 2022 and will go high as \$10,500B in the following three years~\cite{cybersecurity-ventures}. Moreover, current cyberattacks are sophisticated and often consist of combinations of multiple adversarial techniques deployed by cybercrime groups and malware. For example, in Fig~\ref{fig:attack-example}, we show actual attacks demonstrating how adversaries use multiple adversarial techniques. The figure shows that admin@338~\cite{admin338}, a cybercrime group, first uses the command line interface to collect information about existing users and then sends phishing emails to trick users into executing malicious files. Another cybercrime group, APT-C-36~\cite{aptC36} first obfuscates email attachments to bypass malware checks and then sends phishing emails to trick users into executing code via the terminal. Through code execution, attackers send the collected information back to a remote server via file transfer. Adversaries launching attacks through various adversarial techniques are often hard to detect, and thus organizations face difficulty in defending their systems~\cite{helpnet-security}.

Attackers can breach defense mechanisms through multiple attack vectors, and thus defending against attacks requires an understanding of how adversaries deploy techniques in combination to compromise defenses~\cite{fireye}. Hence, cybersecurity analysts identify specific adversarial techniques in cyberattacks and publish press articles and technical reports on the usage. The MITRE ATT\&CK framework~\cite{mitreAttack} maintains a catalog of the adversarial techniques used by cybercrime groups and in malware activities documented in these reports. The catalog contains information on what techniques adversaries used in cyberattacks and thus, reflects adversaries breaching systems from the aspect of adversarial techniques used in conjunction. Thus, the ATT\&CK catalog enables researchers to analyze various aspects of cyberattacks through atomic adversarial behaviors based on the cataloged techniques. For example, in the literature, researchers have utilized the MITRE ATT\&CK framework to perform cyberthreat attribution~\cite{abuAttributionCyberattackUsing2020, kimAutomaticallyAttributingMobile2021, weiDeepHunterGraphNeural2021a}, malware profiling~\cite{fairbanksIdentifyingATTCK2021, huangOpenSourceIntelligence2021}, and tracking the provenance of attack indicators along with utilized/impacted resources~\cite{hassanTacticalProvenanceAnalysis2020}. 

Although MITRE ATT\&CK enables researchers to correlate malware traces and intrusion alerts to adversarial techniques~\cite{beradyTTPIoCAdvanced2021, milajerdiHOLMESRealTimeAPT2019}, thwarting cyberattacks requires the understanding of the complex relationship among adversarial techniques. The pyramid of pain~\cite{pyramidOfPain}, a conceptual model of thwarting adversaries by responding to detected indicators, also emphasizes that responding to adversarial techniques is the most effective albeit the toughest way to prevent cyberattacks. Thus, the ATT\&CK catalog of adversarial techniques used together in cybercrime groups and malware can aid practitioners in capturing the relationship among techniques. Practitioners can also look for evidence of other potential adversarial techniques in the victim environment. Practitioners can also identify to what extent security enforcement is performing and prioritize what ATT\&CK techniques they should mitigate first to facilitate mitigating other ATT\&CK techniques. Practitioners can devise proactive detection and mitigation strategies to improve security practices. \textit{The goal of this research is to aid cybersecurity practitioners and researchers in choosing detection and mitigation strategies through co-occurrence analysis of adversarial techniques reported in MITRE ATT\&CK.} We investigate the following research questions (RQs): 

\begin{itemize}

    \item \textbf{RQ1:} What are the top techniques used by adversaries? What are the top co-occurring techniques reported in adversary activities? 
    
    \item \textbf{RQ2:} Given the occurrence of a technique, how can we predict what other associated techniques adversaries can use? What adversary behavior regarding technique usage do the co-occurrences indicate? 
    
\end{itemize}

To answer the research questions, we collect adversarial technique usage data by cybercrime groups and malware from the MITRE ATT\&CK framework. To answer RQ1, we use frequent itemset mining to identify the following used by adversaries: (a) a set of top adversarial techniques; (b) a set of top co-occurring adversarial techniques. To answer RQ2, we first use association rule mining to identify a set of probabilistic rules of co-occurring techniques. We then build a co-occurrence network from the obtained rules and apply three centrality measures to identify adversary behaviors through technique co-occurrences. Overall, we list our contribution:
\begin{enumerate}
    \item Identification of top adversarial techniques used in conjunction by adversaries, thus benefiting cybersecurity researchers and practitioners in understanding cyberattacks from the perspective of leveraging techniques; 
    \item Identification of co-occurrence rules among adversarial techniques that can aid practitioners in threat hunting through predicting potential technique(s) in use based upon the evidence of a given technique; 
    \item Identification of detection, and mitigation strategies, and adversary behaviors from techniques co-occurrences that can aid practitioners in deriving mitigation strategies; 
    \item Practitioners can use the methodology and findings of the paper as a starting point to investigate adversarial techniques in their environment, derive mitigation strategies, and improve security practices. The dataset and analysis script are made available (see~\cref{dataset}) to replicate the study on future ATT\&CK versions and any other co-occurrence data on the usage of adversarial techniques. 
\end{enumerate}

The rest of the paper is organized below. In~\cref{sec:concept}, we discuss a few key concepts. In~\cref{sec:method}, we discuss the methodology of our work. In~\cref{sec:resultRQ1}, and ~\cref{sec:resultRQ2}, we discuss our findings. We~\cref{sec:discussion}, we reflect on observations from the findings and future research paths. In~\cref{sec:threats}, we discuss several study limitations. In~\cref{sec:related}, we discuss related works and in~\cref{sec:conclusion}, we conclude the paper. 

\section{Key Concepts}
\label{sec:concept}
In this section, we discuss concepts related to the methodology and findings of this study. 

\subsection{Tactics, techniques and procedures (TTPs)} 
Threat actors utilize a plethora of tactics, techniques, and procedures (TTPs) to compromise the security of target organizations or systems. \textit{Tactic} refers to adversary’s tactical goal for performing an action, \textit{Techniques} refers to \textit{how} an adversary achieves a tactical goal by performing an action~\cite{mitreAttack, attack-design}. \textit{Procedures} refer to the specific method adversaries have used for implementing techniques to achieve a particular tactic~\cite{attack-design}. TTPs are used to profile and analyze the lifecycle and behavior of adversaries launching attacks on a targeted system~\cite{nist-glossary}. For example, \textit{privilege escalation}~\cite{TA0004} is a tactic for gaining elevated permission on a system. An example technique of privilege escalation is \textit{access token manipulation}~\cite{T1134}. Thus, an attacker can gain elevated privilege in a system by tampering with the access token to bypass the access control mechanism. An example procedure for this tactic and procedure is the FIN6 cybercrime group~\cite{fin6} \textit{manipulating an access token} by \textit{using Metasploit's named-pipe impersonation}~\cite{metasploit}.

\subsection{MITRE ATT\&CK}
The MITRE~\cite{mitre} organization introduced the ATT\&CK~\cite{mitreAttack} framework in 2013, derived from real-world observations of adversarial TTPs deployed by cybercrime groups and malware. ATT\&CK catalogs an enumeration of tactics. Each tactic has an enumeration of corresponding techniques. Each technique has an enumeration of corresponding procedure(s). The procedures are collected from online articles and technical reports describing cyberattack incidents. ATT\&CK cites each of the articles and reports their corresponding TTPs. ATT\&CK catalogs and annually updates the sets of TTPs observed in \textit{cybercrime groups}~\cite{attack-groups-def} (i.e., cyber-criminals or malicious campaigns tracked by a common name in the security community) and the \textit{malware}~\cite{attack-malware-def} (i.e., software/tools, scripts, executable used for malicious purposes). We use the enterprise ATT\&CK Version 10 in this study, and this version enumerates 14 tactics and 188 adversarial techniques. 


\subsection{Frequent Itemset Mining}
\label{key_concept_fim}
Frequent Itemset Mining (FIM) refers to the extraction of frequently occurring items, events, and patterns from data~\cite{luna2019frequent}. Hence, FIM can extract techniques that are used frequently by adversaries. For example, in Fig.~\ref{fig:attack-example}, we observe four adversaries using a collection of techniques where the four adversaries use eight techniques in total. Throughout the study, we refer to the collection of techniques used by an adversary as \textit{technique-set}, and hence, Fig.~\ref{fig:attack-example} contains four \textit{technique-sets} used by four adversaries. 

In the scope of FIM, each of the techniques are \textit{items}, and each of the \textit{technique-sets} are \textit{itemset}. FIM takes one input called \textit{minimum support (minSup)} and returns \textit{frequent itemsets} referring to the \textit{itemsets} appearing in at least \textit{minSup}\% of all \textit{itemsets} present in the data. \textit{Frequent itemsets} consist of sets of items (i.e. techniques) and each of the sets can contain a single item (i.e. technique) or multiple items (i.e. collection of techniques). A set containing a single item represents a technique that is used by \textit{minSup}\% of adversaries. A set containing multiple items represents co-occurring techniques used by \textit{minSup}\% of adversaries in the scope of an attack. 

We provide examples from Fig.~\ref{fig:attack-example}. We assume \textit{minSup} is 0.25 which denotes all \textit{frequent itemsets} returned by FIM must appear in at least 1 out of 4 \textit{itemsets}. One such \textit{frequent itemsets} is $\{OB\}$ as the technique OB with $support = 0.5$ (i.e. OB appears in 2 out of 4 \textit{itemsets}) satisfies the \textit{minSup}. Similarly, another \textit{frequent itemsets} is $\{CS\}$ as the technique CS with $support = 0.75$ (i.e. CS appears in 3 out of 4 \textit{itemsets}) satisfies the \textit{minSup}. Note that these two examples show that each of the two \textit{frequent itemsets} contains only a single item, and we refer \textit{frequent itemsets} containing only a single item as \textit{frequent individual technique}. Hence, OB or CS are examples of \textit{frequent individual techniques}. However, FIM also returns \textit{frequent itemsets} containing multiple items such as the following two sets: $\{OB,IT\}$, and $\{PH,UE\}$. We see both techniques, OB and IT, are used together by two adversaries; hence, both appear at 2 out of 4 itemsets satisfying \textit{minSup}. Similarly, we also see both techniques: PH and UE are used together by three adversaries, and hence, both techniques appear at 3 out of 4 itemsets satisfying \textit{minSup}. We refer \textit{frequent itemsets} containing multiple items as \textit{frequent co-occurring techniques}. Hence, $\{OB,IT\}$, and $\{PH,UE\}$ are examples of \textit{frequent co-occurring techniques}.

\subsection{Association Rule Mining}
\label{sec:concepts:amr}
\textit{Frequent co-occurring techniques} obtained from frequent itemset mining indicate potential associations among the techniques. For example, we obtain a \textit{frequent co-occurring techniques}: $\{PH,UE\}$ (see \cref{key_concept_fim}) which indicates that a potential association (such as correlation, causation, relation) exists between the two techniques. Hence, we can identify an association rule $PH \implies UE$, which indicates that if PH appears in an \textit{itemset}, then UE also appears in the same \textit{itemset}. Association Rule Mining (ARM) refers to the extraction of association rules from \textit{itemsets}. As in this study, we are investigating techniques occurring together in a single cybercrime group or malware activity. We refer to association rules as \textit{co-occurrence rules} among adversarial techniques. Hence, $PH \implies UE$ is an example of \textit{co-occurrence rules}. 

A \textit{co-occurrence rule} looks identical to an \textit{if-then} expression. For example, the example \textit{co-occurrence} rule: $PH \implies UE$ denotes that \textit{if} PH occurs, \textit{then} UE also occurs. We refer to the \textit{if} portion of the rule as an \textit{antecedent} (e.g. PH), and the \textit{then} portion of the rule as an \textit{consequent} (e.g. UE). ARM takes two input to extract \textit{co-occurrence} rules: \textit{minimum support (minSup)}, and \textit{minimum confidence (minConf)}. \textit{minSup} denotes that the \textit{co-occurrence rule} materializes in at least \textit{minSup}\% of \textit{itemsets}. \textit{minConf} denotes that, in a \textit{co-occurrence rule}, given that \textit{antecedent} appears in an itemset, \textit{consequent} has a \textit{minConf} probability of appearing in the \textit{itemset}. 

We provide examples from Fig.~\ref{fig:attack-example} where we assume, ARM is run with $minSup = 0.5$, and $minConf = 0.5$. We identify a \textit{co-occurrence rule}: $CS \implies OB$ having $support = 0.5$ and $confidence = 0.66$. We see: (a) both CS and OB appear in 2 out 4 itemsets; (b) CS appears in 3 itemsets and OB appears in 2 out of the 3 itemsets. Note that, in this rule, both \textit{antecedent} and \textit{consequent} contain only one technique each. We refer to such rule as \textit{simple co-occurrence rule}. From Fig.~\ref{fig:attack-example}, we identify another rule: $PH \land UE \implies CS$ having $support = 0.5$ and $confidence = 0.66$. Note that, in this rule, \textit{antecedent} contains two techniques. We refer to \textit{co-occurrence rule} where either \textit{antecedent} or \textit{consequent} contain more than one technique as \textit{compound co-occurrence rule}. 

\begin{figure}
    \centering
    \includegraphics[width=\columnwidth]{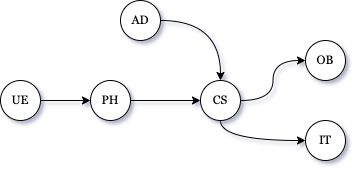}
    \caption{Example of co-occurrence network derived from Fig~\ref{fig:attack-example}}
    \label{fig:network-example}
\end{figure}

\subsection{Co-occurrence network}
A co-occurrence network refers to the graph representing relations between items appearing in the same dataset~\cite{segev2021semantic}. In this study, we use co-occurrence network to represent \textit{co-occurrence rules} among the adversarial techniques. The co-occurrence network is a directed graph where each node represents a technique and each directed edge represents a \textit{simple co-occurrence rule}. We demonstrate how we construct a co-occurrence network obtained from \textit{co-occurrence rules} based on Fig.~\ref{fig:attack-example}. Assume these five following \textit{simple co-occurrence rules} are obtained: $UE \implies PH$, $PH \implies CS$, $AD \implies CS$, $CS \implies IT$, and $CS \implies OB$. Five techniques exist in these five rules. We represent these five techniques by five nodes and then, we represent \textit{co-occurrence rules} by directed edge from \textit{antecedent} techniques to \textit{consequent} technique. Thus, we obtain a co-occurrence network shown in Fig.~\ref{fig:network-example}. 

Note that, while each edge in the network represents a \textit{simple co-occurrence rule}, a path in the network represents a \textit{chained co-occurrence rule} referring to two or more rules being chained together. For example, a path: (PH, CS, OB) represents a \textit{chained co-occurrence rule}: $PH \implies CS \implies OB$. The path denotes the following: (a) given that PH occurs, CS is likely to occur; (b) given that both PH and CS occur, OB is likely to occur. In the \textit{chained co-occurrence rule}, CS works as an antecedent for OB and consequent for PH. Hence, we refer to the intermediate nodes in a path as (e.g. CS) as \textit{intermediate antecedent} or \textit{intermediate consequent}. We use these two terms interchangeably throughout the paper. Also, note that while PH can directly imply the probability of occurrence of CS. However, PH cannot directly imply the probability of occurrence of OB because, in the given rule, PH can only imply the occurrence of OB, given that CS also occurs. Thus, in a \textit{chained co-occurrence rule}, the probability of a consequent depends on both the probability of its \textit{intermediate antecedent}, and \textit{antecedent}. Hence, the longer a path is in the network, the extent of an \textit{antecedent}, implying the probability of the occurrence of the \textit{consequent} is weaker. A shortest path between an \textit{antecedent} and a \textit{consequent} reflects the strongest implication of occurrence of the \textit{consequent} given that the \textit{antecedent} occurs.  

\subsection{Centrality Measures}
In the context of graphs or networks, centrality measures refer to the computation of how \textit{important} a node is compared to other nodes~\cite{kempf-leonard_network_2005}. The notion of \textit{importance} depends on the context and structure of the graph or network. Hence, in a co-occurrence network, centrality measures can indicate the importance of adversarial techniques in the context of how they imply the occurrence of one another. In this study, we apply the three centrality measures. We discuss the measures from the perspective of a hypothetical technique $te$ in Table~\ref{tab:cent-def}.

\begin{table*}[]
    \centering
    \caption{Centrality measures from the perspective of a technique: $te$ compared to other techniques in the network}
    \begin{tabular}{p{25mm}p{25mm}p{75mm}p{4cm}}
    \toprule
    \textbf{Centrality} & \textbf{Definition} & \textbf{Explanation} & \textbf{Implication} \\ \midrule
    
    In-degree centrality~\cite{golbeck_chapter_2013} & Number of incoming edges & Higher value indicates $te$ is common consequent in relatively higher number of rules. E.g., in Fig.~\ref{fig:network-example}, CS has the highest in-degree centrality score of the other four. CS has two incoming edges, while the remaining four have one incoming edge.  & A relatively high number of techniques imply the occurrence of $te$ \\ \midrule
    
    Out-degree centrality~\cite{golbeck_chapter_2013} & Number of outgoing edges & Higher value indicates $te$ is common antecedent in relatively higher number of rules. E.g., in Fig.~\ref{fig:network-example}, CS has the highest out-degree centrality score of the other four. CS has two outgoing edges while the remaining four has one outgoing edge & $te$ implies the occurrence of a relatively higher number of other techniques \\ \midrule
    
    Betweenness centrality~\cite{freeman1977set} & Number of shortest paths between any two techniques in the network where the paths go through $te$ & Higher value indicates $te$ acts as an \textit{intermediate antecedent} or \textit{consequent} in a relatively higher number of the shortest chained rules between any two other technique. E.g. in Fig.~\ref{fig:network-example}, CS is an \textit{intermediate antecedent/consequent} of 6 shortest paths between each of: (UE, OB), (UE, IT), (PH, OB), (PH, IT), (AD, OB), and (AD, IT) & Any two techniques' co-occurrence depends on the occurrence of $te$ more than that of any other techniques \\

    \bottomrule
    \end{tabular}
    \label{tab:cent-def}
\end{table*}

\section{Methodology}
\label{sec:method}
\begin{figure}
    \centering
    \includegraphics[width=\columnwidth]{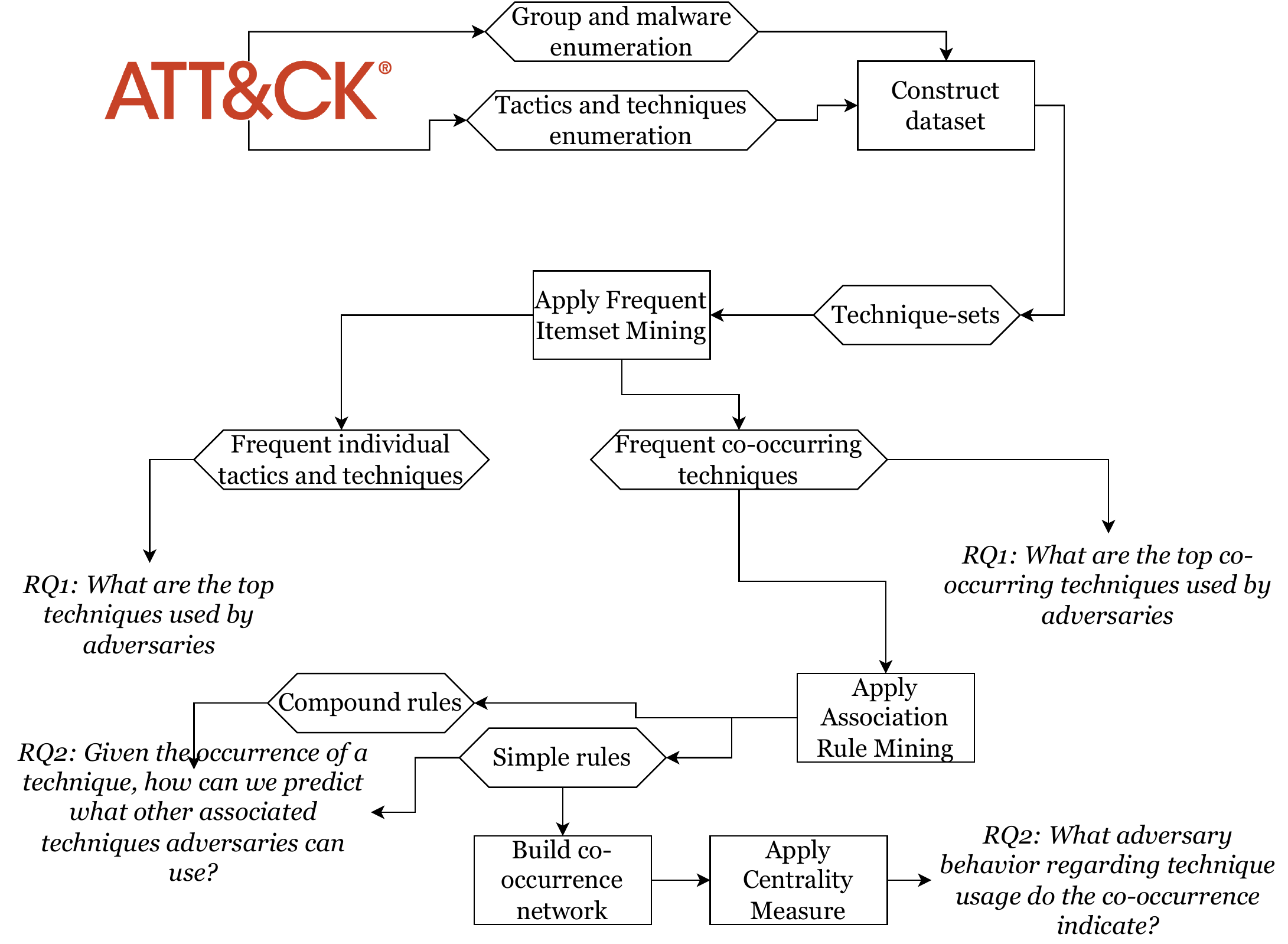}
    \caption{An Overview of the Methodology}
    \label{fig:methodology}
\end{figure}

We discuss the methodology in this section. An overview of the methodology appears in Fig.~\ref{fig:methodology}. The rectangle boxes represent steps in the method. The hexagonal boxes represent information that is the input/output of each step. We also use arrows from steps to RQs to show which step is used to draw observations to answer the RQs.

\subsection{Construct Dataset}
\begin{figure}
    \centering
\includegraphics[width=\columnwidth]{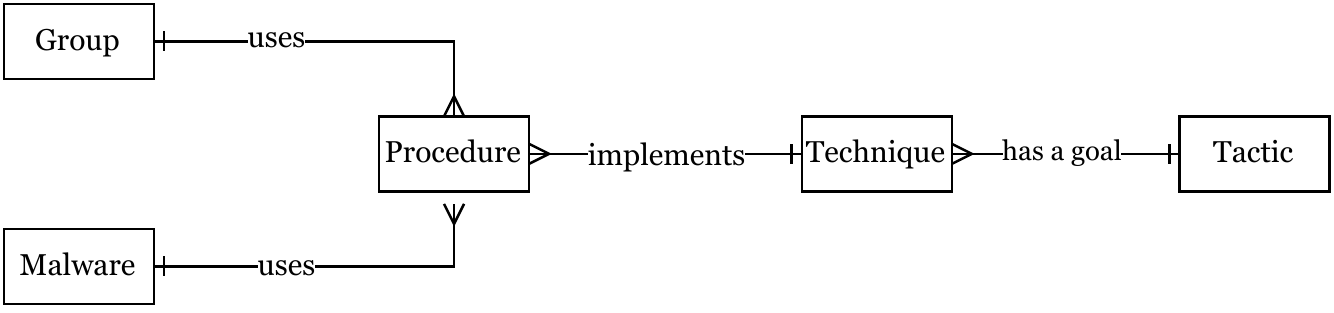}
    \caption{The Dataset Schema}
    \label{fig:schema}
\end{figure}

MITRE ATT\&CK catalogs a collection of cybercrime groups and malware whose activities are reported in publicly-accessible documents. MITRE ATT\&CK catalogs the procedures and the corresponding adversarial technique(s) and tactics used by each group or malware during cyberattacks. We obtain the catalog of groups and malware along with the associated techniques from the MITRE ATT\&CK website~\cite{db-attack-groups, db-attack-malware}. We next obtain the catalog of tactics and techniques from MITRE ATT\&CK website~\cite{dbattacktactics, db-attack-techniques}. We then combine the acquired data to build the dataset according to the schema shown in Fig~\ref{fig:schema}. The schema shows: (a) each group or malware uses a collection of techniques where the group or malware implements techniques through procedures; (b) the adversary can use multiple techniques to accomplish a single tactic. 

\begin{figure}
    \centering
    \includegraphics[width=\columnwidth]{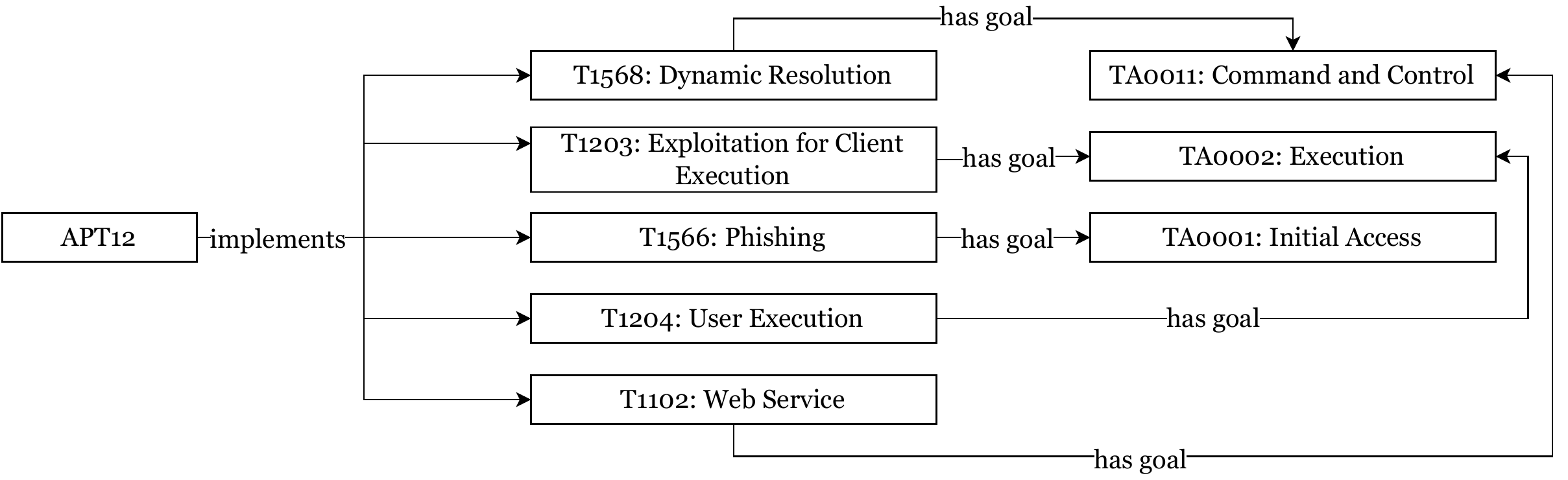}
    \caption{Dataset Example: APT12 group~\cite{apt12}}
    \label{fig:schema-example}
\end{figure}

Fig.~\ref{fig:schema-example} provides a real example according to the schema. In the example, we show the reported techniques of a cybercrime group named APT12~\cite{apt12} who targeted numerous government organizations, technology companies, and media outlets. The group has been reported to use five adversarial techniques to gain three tactical goals: (a) use of \textit{T1566: Phishing}~\cite{T1566} technique to achieve the tactical goal \textit{TA0001: Initial access}~\cite{TA0001}; (b) use of \textit{T1203: Exploitation for client execution}~\cite{T1203} and \textit{T1204: User execution}~\cite{T1204} technique to achieve the tactical goal \textit{TA0002: Execution}~\cite{TA0002}; (c) use of \textit{T1568: Dynamic resolution}~\cite{T1568} and \textit{T1102: Web service}~\cite{T1102} techniques to achieve the tactical goal \textit{TA0011: Command and control}~\cite{TA0011}. Thus, the dataset contains a collection of groups and malware. We refer to the groups and malware as adversary entities. Each entity contains a collection of techniques reported to be used by the corresponding adversary entity. We refer to the collection of techniques used by an adversary entity as \textit{technique-set} (\cref{key_concept_fim}). We then keep only the \textit{technique-sets} if the \textit{technique-set} contains at least three techniques. 

\subsection{Apply frequent itemset mining}
\label{sec:method-fim}
We apply frequent itemset mining on technique-sets to identify: (a) frequent individual techniques and (b) frequent co-occurring techniques. We set $minSup = 0.10$, and thus the mining returns frequent individual techniques and frequent co-occurring techniques appearing in at least 10\% of all the technique-sets. We obtain the \textit{tactic-set} from each of the corresponding \textit{technique-sets} and then apply frequent itemset mining with $minSup = 0.10$ to find what set of tactics are aimed by adversaries from their corresponding used techniques. We use \textit{mlxtend}~\cite{mlxtend} package to perform the mining. We answer the \textit{RQ1} based upon our observation on the obtained frequent individual tactics and techniques in this step. 

\subsection{Apply association rule mining}
\label{sec:method-arm}
We apply association rule mining to the obtained frequent co-occurring techniques to identify simple and compound co-occurrence rules among the techniques. We set $minSup = 0.10$, and $minConf = 0.15$. Thus the mining returns rules that materialize in at least 10\% of all the technique-sets, and given that the antecedent technique occurs, the consequent technique occurs with at least 0.15 probability. We use \textit{mlxtend} package to perform the mining. We answer the \textit{RQ2: Given the occurrence of a technique, how can we predict what other associated techniques adversaries can use?} based upon our observation on obtained rules in this step.

\begin{table}[t]
    \centering
    \caption{Summary of the dataset}
    \begin{tabular}{lrrrrrrr}
    \toprule
        \multirow{2}{*}{\textbf{Adversary}} & \multirow{2}{*}{\textbf{Count}} & \multicolumn{3}{c}{\textbf{Technique}} & \multicolumn{3}{c}{\textbf{Tactic}} \\
        {} & {} & \textbf{Avg (Med)} & \textbf{Min} & \textbf{Max} & \textbf{Avg (Med)} & \textbf{Min} & \textbf{Max} \\ \midrule
        Groups & 115 & 19.0 (13) & 3 & 64 & 7.1 (7) & 2 & 14 \\ 
        Malware & 484 & 12.3 (11) & 3 & 56 & 5.0 (5) & 1 & 11 \\ \midrule
        Total & 599 & 13.6 (11) & 3 & 64 & 5.5 (5) & 1 & 14 \\ 
    \bottomrule
    \end{tabular}
    \label{tab:dataset-summary}
\end{table}

\subsection{Build co-occurrence network}
\label{sec:method-coc}
After applying association rule mining, we obtain a set of simple and compound co-occurrence rules(~\cref{sec:concepts:amr}). Each simple co-occurrence rule contains antecedent and consequent techniques. We create a co-occurrence network from the set of obtained simple co-occurrence rules where the confidence value of the rule is at least greater than or equal to the median confidence value of all simple co-occurrence rules. The co-occurrence network is a directed graph where each node represents a technique. We create directed edges between two techniques such that the direction goes from antecedent to consequent techniques. For each node, we record the tactic of the corresponding technique and the number of times the technique appears in the dataset. For each edge, we record the number of times the same adversary entity uses both the antecedent and consequent techniques.

\subsection{Apply centrality measures}
\label{sec:method-cent}
The constructed co-occurrence network reflects the probabilistic ties among the co-occurrence of adversarial techniques. Hence, in the co-occurrence network, centrality measures can indicate the importance of adversarial techniques in how their co-occurrence is probabilistic-ally tied to one another. In this study, we apply three centrality measures discussed in Table~\ref{tab:cent-def}. We use \textit{Networkx}~\cite{networkx} package to compute the centrality measures, and we use the default parameters provided by Networkx while computing each measure. We answer the \textit{RQ2: What adversary behavior regarding technique usage do the co-occurrence indicate?} based upon our observation from the computed measures in this step. 

\section{The Constructed Dataset}
After constructing the dataset according to the schema, we obtain 669 technique-sets used by 125 groups and 544 malware. We then drop the technique-sets where each of the dropped technique-sets contains less than three techniques. Thus, we obtain 599 technique-sets used by 115 groups and 484 malware.  We present a summary of the dataset in Table~\ref{tab:dataset-summary}. Overall, the 599 technique-sets contain 172 of the 188 techniques, and the 172 techniques cover all 14 of the tactics cataloged in MITRE ATT\&CK.

\section{Findings on RQ1}
\label{sec:resultRQ1}
In this section, we discuss our findings on frequent individual tactics, frequent individual techniques, and co-occurring techniques to answer RQ1. 

\begin{table}[]
    \centering
    \caption{Top ten individual techniques used by adversaries}
    \begin{tabular}{lrr}
         \toprule 
         \textbf{Technique} & \textbf{Tactic} & \textbf{Support} \\ \midrule
         T1059: Command and Scripting Interpreter~\cite{T1059} & EX & 0.62 \\ 
         T1105: Ingress Tool Transfer~\cite{T1105} & CC & 0.56 \\ 
         T1027: Obfuscated Files or Information~\cite{T1027} & DE & 0.51 \\ 
         T1071: Application Layer Protocol~\cite{T1071} & CC & 0.47 \\ 
         T1082: System Information Discovery~\cite{T1082} & DC & 0.47 \\ 
         T1083: File and Directory Discovery~\cite{T1083} & DC & 0.40 \\ 
         T1070: Indicator Removal on Hosts~\cite{T1070} & DE & 0.39 \\ 
         T1547: Boot or Logon Autostart Execution~\cite{T1547} & PE & 0.36 \\ 
         T1057: Process Discovery~\cite{T1057} & DC & 0.34 \\ 
         T1016: System Network Configuration Discovery~\cite{T1016} & DC & 0.32 \\ 
         \midrule
         \multicolumn{2}{l}{Total techniques on $support > 0.1$} & 44 \\
         \multicolumn{2}{l}{Mean support of techniques on $support > 0.1$} & 0.23 \\
         \multicolumn{2}{l}{Median support of techniques on $support > 0.1$} & 0.18 \\
         \multicolumn{2}{l}{Total techniques on $0.05 <  support < 0.1$} & 24 \\
         \multicolumn{2}{l}{Total techniques on $0.01 <  support < 0.05$} & 63 \\
         \multicolumn{2}{l}{Total techniques on $support < 0.01$} & 41 \\
         \multicolumn{2}{l}{Total techniques} & 172 \\ \midrule
         \multicolumn{3}{p{8cm}}{EX: TA0002 - Execution~\cite{TA0002}, CC: TA0011 - Command and control~\cite{TA0011}, DE: TA0005 - Defense evasion~\cite{TA0005}, DC: TA0007 - Discovery~\cite{TA0007}, PS: TA0003- Persistence~\cite{TA0003}} \\
         \bottomrule
    \end{tabular}
    \label{tab:top-ten-technique-support}
\end{table}

\subsection{Frequent individual techniques used by adversaries}
We obtain a set of frequent individual techniques from applying frequent itemset mining (\cref{sec:method-fim}). In Table~\ref{tab:top-ten-technique-support}, we show the Top 10 techniques used by the adversaries. \textit{Support} column in the table denotes the percentage of adversary entities (i.e., cybercrime groups and malware) using the technique. For example, the table shows the top technique is \textit{T1059: Command and Scripting Interpreter} which is used by 62\% of adversary entities. The table shows, 62\%, 56\%, and 51\% of the adversary entities use the first three techniques, respectively. However, the bottom three techniques have relatively high support with use, respectively, by 36\%, 34\%, and 32\%, of the adversary entities. The total number of techniques used by at least 10\% of adversary entities is 44. However, the number of techniques in the dataset is 172. The observation suggests that 26\% of the techniques in the dataset are used by at least 10\% of adversary entities. We also identify 104 techniques appearing in less than 5\% of adversary entities, and these 104 techniques constitute 60\% of all techniques in the dataset. While adversaries use these Top 10 techniques for malicious purposes, the techniques are an abuse of legitimate system functions. Thus, detecting these techniques is challenging, and filtering malicious activities from benign system activities becomes difficult. We manually checked the procedures of the top three techniques from the ATT\&CK and identified the following: (a) \textit{T1059} is mainly used for downloading and executing malicious scripts; (b) \textit{T1105} is mainly used for downloading scripts and tools; and (c) \textit{T1082} is mainly used for encrypting/encoding malicious payload and commands. From the observation, we identify the following detection and mitigation strategies (DS and MS, respectively), below.

\noindent \fbox{\begin{minipage}{24.5em}
\ding{43} \textbf{DS-1}: Approximately half of the adversaries have three common features: command executing, downloading of files, and encrypted/encoded files or payloads. Thus, organizations can detect a relatively large number of malicious activities by detecting indicators of these three features.  

\ding{43} \textbf{MS-1}: Enforcing mitigation for all 188 ATT\&CK techniques may be infeasible for organizations with limited resources. However, organizations can achieve reasonable protection against one-third of the adversaries by mitigating a small (n=10) number of techniques. 
\end{minipage}}

As shown in Table~\ref{tab:top-ten-technique-support}, the Top 10 techniques are from five tactics. \textit{TA0007: Discovery} has four techniques. Both \textit{TA0011: Command and Control} and \textit{TA0005: Defense Evasion} have two techniques each. \textit{TA0002: Execution} and \textit{TA0003: Persistence} have one technique each. The top three techniques are from \textit{TA0002: Execution}, \textit{TA0011: Command and Control}, and \textit{TA0005: Defense evasion} respectively. The observation provides a bigger picture of generic intrusion activity. Adversaries primarily figure out the victim system by obtaining information, hiding the evidence of intrusion, remaining in the system by attaching malicious programs to periodic system functions, and executing malicious tasks through commands. By definition, techniques from \textit{TA0003: Persistence}, and \textit{TA0005: Defense Evasion} are challenging to detect, and techniques from \textit{TA0002: Execution} indicate a potential breach. Meanwhile, techniques from \textit{TA0007: Discovery} indicate potentially an early phase of a breach where the adversary is figuring out its follow-on course of actions. However, ATT\&CK reports that no mitigation exists for the four \textit{TA0007: Discovery} related techniques, and \textit{T1547: Boot or Logon autostart execution}~\cite{T1082, T1547, T1057, T1016, T1083}. 

\noindent \fbox{\begin{minipage}{24.5em}
\textbf{DS-2}: Five of the Top 10 techniques: \textit{T1082, T1547, T1057, T1016, T1083} do not have any mitigation. Hence, detection of these five techniques is of topmost priority to make an early intervention to stop system breaches.
\end{minipage}}

\begin{table*}[]
    \centering
    \caption{Techniques used by at least ten percent of adversaries from each ATT\&CK tactics}
    \begin{tabular}{lrrrrrrrl}
    \toprule
    \textbf{Tactic} & \textbf{Support} & \textbf{Count} & \textbf{Min} & \textbf{Avg} & \textbf{Med} & \textbf{Std.} & \textbf{Max} & \textbf{Most used technique} \\ \midrule
    
     TA0005: Defense Evasion~\cite{TA0005}      &      0.85 &      12 &   0.1  &   0.22 &   0.18 &    0.13 &   0.51 & T1027: Obfuscated Files or Information~\cite{T1027}   \\
     
     TA0011: Command and Control~\cite{TA0011}  &      0.79 &       6 &   0.12 &   0.28 &   0.2  &    0.19 &   0.56 & T1105: Ingress Tool Transfer~\cite{T1105}             \\
     
     TA0002: Execution~\cite{TA0002}            &      0.76 &       5 &   0.14 &   0.27 &   0.21 &    0.2  &   0.62 & T1059: Command and Scripting Interpreter~\cite{T1059} \\
     
     TA0007: Discovery~\cite{TA0007}            &      0.75 &       9 &   0.11 &   0.26 &   0.25 &    0.13 &   0.47 & T1082: System Information Discovery~\cite{T1082}         \\
     
     TA0003: Persistence~\cite{TA0003}          &      0.53 &       2 &   0.2  &   0.28 &   0.28 &    0.11 &   0.35 & T1547: Boot or Logon Autostart Execution~\cite{T1547}    \\
     
     TA0009: Collection~\cite{TA0009}           &      0.51 &       5 &   0.13 &   0.18 &   0.19 &    0.03 &   0.22 & T1056: Input Capture~\cite{T1056}                     \\
     
     TA0006: Credential Access~\cite{TA0006}    &      0.28 &       2 &   0.13 &   0.13 &   0.13 &    -    &   0.13 & T1555: Credentials from Password Stores~\cite{T1555}  \\
     
     TA0001: Initial Access~\cite{TA0001}       &      0.25 &       1 &   0.18 &   0.18 &   0.18 &    -    &   0.18 & T1566: Phishing~\cite{T1566}                          \\
     
     TA0010: Exfiltration~\cite{TA0010}         &      0.25 &       1 &   0.16 &   0.16 &   0.16 &    -    &   0.16 & T1041: Exfiltration Over C2 Channel~\cite{T1041}            \\
     
     TA0008: Lateral Movement~\cite{TA0008}     &      0.16 &       1 &   0.12 &   0.12 &   0.12 &    -    &   0.12 & T1021: Remote Services~\cite{T1021}                   \\
     
     TA0042: Resource Development~\cite{TA0042}     &      0.10 &       0 &   - &   - &   - &    -    &   - & -                   \\
     
     TA0004: Privilege Escalation~\cite{TA0004}     &      0.05 &       0 &   - &   - &   - &    -    &   - & -                   \\
     
     TA0043: Reconnaissance~\cite{TA0043}     &      0.03 &       0 &   - &   - &   - &    -    &   - & -                   \\
     
     TA0040: Impact~\cite{TA0040}     &      0.02 &       0 &   - &   - &   - &    -    &   - & -                   \\
     
    \bottomrule
    \end{tabular}
    \label{tab:tactics-support}
\end{table*}


We investigate to what extent adversary entities use techniques from each of the 14 ATT\&CK tactics. We report the findings  Table~\ref{tab:tactics-support}. We show the following for each tactic: (a) support denoting the percentage of entities using at least one technique; (b) the total number of techniques from the corresponding tactics having at least 0.1 support; (c) minimum, mean, median, standard deviation, and a maximum of support values of the techniques; and (d) the technique having maximum support value. The first row in the table denotes the following: (a) 85\% of the entities use at least one technique from \textit{TA0005: Defense Evasion} tactic; (b) 12 techniques from the tactic have minimum support of 0.1 (c) among the support values of 12 techniques, minimum and maximum values are respectively 0.1 and 0.51; (d) \textit{T1027: Obfuscated Files or Information} technique has the highest support value. 

We do not find any technique from \textit{TA0043: Reconnaissance}, \textit{TA0042: Resource Development}, \textit{TA0004: Privilege Escalation}, and \textit{TA0040: Impact} tactics with support greater than 0.10. No technique from \textit{TA0043: Reconnaissance}, \textit{TA0042: Resource Development} may indicate, how adversaries initially plan their operation at an early phase of an attack is hard to know for the organizations. No technique from \textit{TA0040: Impact} may indicate that adversaries rather focus on remaining hidden and persistent instead of using a hit-and-run approach. However, no technique from the \textit{TA0004: Privilege escalation} tactic may indicate adversaries circumvent \textit{TA0004: Privilege escalation} through performing techniques from \textit{TA0001: Initial access}, \textit{TA0006: Credential access}, and \textit{TA0009: Collection}. The tactics may aid an adversary in obtaining required information for the escalating privilege by gaining credentials. Another reason could be that privilege escalation attempts are not detected in the same proportions as that of other tactics.

%

Table~\ref{tab:tactics-support} shows that adversaries primarily use techniques from the top four tactics: \textit{TA0005: Defense evasion, TA0011: Command and control, TA0002: Execution, and TA0007: Discovery}. 75\% of the adversary entities use at least one technique from each of the four tactics. We also observe that 50\% of the entities use at least one technique each from \textit{TA0003: Persistence}, and \textit{TA0009: Collection}. \textit{TA0005: Defense Evasion}, and \textit{TA0007: Discovery} tactics have the two highest number of techniques: 12 and 9 respectively. The table shows techniques having the highest support from each tactic, and we observe that techniques from the top five tactics in the table are among the Top 10 techniques shown in Table~\ref{tab:top-ten-technique-support}. 

A tactic can have a comparatively lower number of techniques but comparatively higher support, suggesting that adversaries use the techniques from the tactic more often than the techniques from other tactics. For example, the \textit{TA003: Persistence} tactic has only two techniques and has a support value of 0.53. However, the \textit{TA0009: Collection} tactic has five techniques and has a support value of 0.51. The \textit{TA0003: Persistence} tactic also has the highest median and mean value of the support of its corresponding techniques. However, the tactic has only two techniques. In the case of support values of the techniques, we also observe that (a) \textit{TA0009: Collection} has the lowest standard deviation; (b) \textit{TA0001: Initial access} has the highest minimum value. The observation indicates that while adversaries lean toward using \textit{T1566: Phishing} to gain initial access to systems, they deploy multiple techniques of \textit{TA0009: Collection} tactic with similar probabilities. Adversaries use the highest number of techniques for \textit{TA0005: Defense evasion} with 85\% of the entities using at least one technique from this tactic. The observation indicates a primary trait of multi-stage attacks that the adversaries always intend to keep their presence hidden, enabling them to conduct their operation persistent.

\noindent \fbox{\begin{minipage}{24.5em}
\ding{43} \textbf{MS-2}: We identify the highest number of techniques are used by the adversaries to achieve the goal of \textit{TA0005: Defense evasion}. Organizations can prioritize the mitigation of techniques of the tactic so that adversaries cannot remain hidden for a prolonged period of time. 
\end{minipage}}

\begin{table}[]
    \centering
    \caption{Top ten co-occurring techniques used together by adversaries}
    \begin{tabular}{lr}
    \toprule 
    \textbf{Combination of Techniques} & \textbf{Support} \\ \midrule
    T1059/EX \& T1105/CC & 0.40 \\ 
    T1059/EX \& T1027/DE & 0.37 \\
    T1105/CC \& T1027/DE & 0.35 \\ 
    T1059/EX \& T1071/CC & 0.34 \\ 
    T1105/CC \& T1071/CC & 0.33 \\ 
    T1059/EX \& T1082/DC & 0.33 \\ 
    T1105/CC \& T1082/DC & 0.32 \\ 
    T1027/DE \& T1071/CC & 0.30 \\ 
    T1059/EX \& T1070/DE & 0.29 \\ 
    T1027/DE \& T1082/DC & 0.29 \\ 
    \midrule
    \multicolumn{2}{l}{Total sets: 188, Unique Techniques: 33, Mean: 0.16, Median: 0.13 } \\ \midrule
    \multicolumn{2}{p{8cm}}{T1059: Command and Scripting Interpreter~\cite{T1059}, T1105: Ingress Tool Transfer, T1027: Obfuscated Files or Information~\cite{T1027}, T1071: Application Layer Protocol~\cite{T1071}, T1082: System Information Discovery~\cite{T1082}, T1070: Indicator Removal on Hosts~\cite{T1070} } \\ \midrule
    \multicolumn{2}{p{8cm}}{EX: TA0002 - Execution~\cite{TA0002}, CC: TA0011 - Command and control~\cite{TA0011}, DE: TA0005 - Defense evasion~\cite{TA0005}, DC: TA0007 - Discovery~\cite{TA0007}} \\
    \bottomrule
    \end{tabular}
    \label{tab:top-ten-technique-couples}
\end{table}

\subsection{Frequent co-occurring techniques}
Our analysis found 188 sets of frequent co-occurring techniques where each set appears in at least 10\% of the adversary entities. We identify 33 techniques used in these 188 sets. Discussing all 188 co-occurrences is out of scope. Hence, we report the Top 10 co-occurrences having the highest support value in Table~\ref{tab:top-ten-technique-couples}. The corresponding tactic of a technique is reported followed by a forward slash.

We observe six techniques in the Top 10 sets. We also observe that \textit{T1059: Command and Scripting Interpreter} appears in five sets, \textit{T1105: Ingress Tool Transfer}, and \textit{T1027: Obfuscated Files or Information} appear in 4 sets. The top 2 co-occurrences also suggest that \textit{T1059: Command and Scripting Interpreter} appears most with \textit{T1105: Ingress Tool Transfer} and \textit{T1027: Obfuscated Files or Information}. A potential correlation might exist between co-occurring techniques, such as co-occurring techniques that may aid the adversary in exploiting a single attack vector. For example, using \textit{remsec}~\cite{remsec} malware, adversary can \textit{download} malicious tools through \textit{T1105: Ingress tool transfer} using \textit{T1071: Application layer protocol}. Given two techniques, an adversary may also use the first technique as a requirement to use the second technique. For example, \textit{chimera}~\cite{chimera} group applied \textit{T1082: System information discovery} technique by \textit{executing fsutil} command, which is an use of \textit{T1059: Command and scripting interpreter} technique. Thus, co-occurring techniques may help defenders identify other potential correlated techniques, given the identification of a technique. We also observe that the ten co-occurrence are among the four tactics: \textit{TA0002: Execution}, \textit{TA0005: Defense evasion}, \textit{TA0007: Discovery}, and \textit{TA0011: Command and control}. Among 6 out of the 10 co-occurrences, we find: (a) \textit{TA0002: Execution} and \textit{TA0005: Defense evasion} twice; (b) \textit{TA0005: Defense evasion}, and \textit{TA0011: Command and control} twice; (c) \textit{TA0002: Execution} and \textit{TA0011: Command and control} twice. 


\noindent \fbox{\begin{minipage}{24.5em}
\ding{43} \textbf{DS-3}: Adversaries' chosen set of techniques primarily achieve three goals: \textit{TA0002: Execution}, \textit{TA0005: Defense evasion}, and \textit{TA0011: Command and control}. Our results indicate that adversaries will likely leverage techniques from these three tactics to breach a system. Hence, detection from one of these tactics can potentially indicate adversaries using techniques from the other two tactics.
\end{minipage}}

\begin{table*}[]
    \centering
    \caption{Top 10 simple co-occurrence rules along with citations to the definitions of the techniques in ATT\&CK}
    \begin{tabular}{llrr}
    \toprule
        \textbf{SI.} & \textbf{Rules} & \textbf{Supp.} & \textbf{Conf.} \\ \midrule
        1 & T1566: Phishing/IA~\cite{T1566} $\implies$ T1204: User Execution/EX~\cite{T1204} & 0.17 & 0.95 \\
        2 & T1033: System Owner/User Discovery/DC~\cite{T1033} $\implies$ T1082: System Information Discovery/DC~\cite{T1082} & 0.21 & 0.86 \\
        3 & T1132: Data Encoding/CC~\cite{T1132} $\implies$ T1059: Command and Scripting Interpreter/EX~\cite{T1059} & 0.12 & 0.86 \\
        4 & T1053: Scheduled Task or Job/EX~\cite{T1053} $\implies$ T1059: Command and Scripting Interpreter/EX & 0.17 & 0.83 \\
        5 & T1566: Phishing/IA $\implies$ T1059: Command and Scripting Interpreter/EX & 0.15 & 0.82 \\ 
        6 & T1041: Exfiltration Over C2 Channel/EF~\cite{T1041} $\implies$ T1059: Command and Scripting Interpreter/EX  & 0.13 & 0.82 \\
        7 & T1140: Deobfuscate/Decode Files or Information/DE~\cite{T1140} $\implies$ T1027: Obfuscated Files or Information/DE~\cite{T1027} & 0.24 & 0.82 \\
        8 & T1218: System Binary Proxy Execution/EX~\cite{T1218} $\implies$ T1059: Command and Scripting Interpreter/EX & 0.15 & 0.81 \\
        9 & T1204: User Execution/EX $\implies$ T1059: Command and Scripting Interpreter/EX & 0.17 & 0.80 \\
        10 & T1204: User Execution/EX $\implies$ T1566: Phishing/IA & 0.17 & 0.80 \\ \midrule
        \multicolumn{4}{l}{Total: 376, minimum confidence: 0.16, mean confidence: 0.49, median confidence: 0.51} \\ \midrule
        \multicolumn{4}{p{13cm}}{IA: TA0001 - Initial access~\cite{TA0001}, EX: TA0002 - Execution~\cite{TA0002}, DC: TA0007 - Discovery~\cite{TA0007}, CC: TA0011 - Command and control~\cite{TA0002}, EF: TA0010 - Exfiltration~\cite{TA0010}, DE: TA0005 - Defense evasion~\cite{TA0005}} \\
    \bottomrule    
    \end{tabular}
    \label{tab:frequent-rules-simple}
\end{table*}

\section{Findings on RQ2}
\label{sec:resultRQ2}
We discuss our findings on co-occurrence rules and the co-occurrence network to answer the RQ2 in this section.

\subsection{Co-occurrence rules among technique}
\label{sec:rq2:1}
We obtain a set of 376 simple and 4,670 compound co-occurrence rules from applying frequent itemset mining (\cref{sec:method-fim}). We report the Top 10 simple rules in Table~\ref{tab:frequent-rules-simple} and Top 10 compound rules in Table~\ref{tab:frequent-rules-comp}. The corresponding tactic of a technique is reported followed by a forward slash. Both tables show the rules along with the corresponding support and confidence of the rule. The rules are sorted by their corresponding confidence score. For example, in Table~\ref{tab:frequent-rules-simple}, the first rule suggests if \textit{T1566: Phishing} technique is used in $n$ adversarial entities, then we can predict \textit{T1204: User Execution} technique is used in $0.95 \times n$. The table indicates that the first rule materializes in the activities of 17\% of the entities. 
\begin{table*}[]
    \centering
    \caption{Top 10 compound co-occurrence rules}
    \begin{tabular}{lp{12cm}rr}
    \toprule
        \textbf{SI.} & \textbf{Rules} & \textbf{Supp.} & \textbf{Conf.} \\ \midrule
        
        1 & T1105: Ingress Tool Transfer/CC~\cite{T1105} $\land$ T1566: Phishing/IA~\cite{T1566} $\implies$ T1204: User Execution/EX~\cite{T1204} & 0.12 & 0.99 \\
        
        2 & T1105: Ingress Tool Transfer/CC $\land$ T1059: Command and Scripting Interpreter/EX~\cite{T1059} $\land$ T1566: Phishing/IA $\implies$ T1204: User Execution/EX & 0.11 & 0.98 \\
        
        3 & T1105: Ingress Tool Transfer/CC $\land$ T1033: System Owner/User Discovery/DC~\cite{T1033} $\land$ T1057: Process Discovery/DC~\cite{T1057} $\implies$ T1082: System Information Discovery/DC~\cite{T1082} & 0.10 & 0.95 \\
        
        4 & T1027: Obfuscated Files or Information/DE~\cite{T1027} $\land$ T1566: Phishing/IA $\implies$ T1204: User Execution/EX & 0.13 & 0.97 \\
        
        5 & T1027: Obfuscated Files or Information/DE $\land$ T1566: Phishing/IA $\land$ T1059: Command and Scripting Interpreter/EX $\implies$ T1204: User Execution/EX & 0.11 & 0.97 \\ 
        
        6 & T1057: Process Discovery/DC~\cite{T1057} $\land$ T1033: System Owner/User Discovery/DC  $\implies$   T1082: System Information Discovery/DC & 0.13 &   0.96 \\
        
        7 & T1059: Command and Scripting Interpreter/EX $\land$ T1566: Phishing/IA  $\implies$   T1204: User Execution/EX & 0.14  &  0.95 \\
        
        8 & T1033: System Owner/User Discovery/DC $\land$ T1071: Application Layer Protocol/CC~\cite{T1071} $\land$ T1016: System Network Configuration Discovery/DC~\cite{T1016} $\implies$   T1082: System Information Discovery/DC & 0.10 &    0.95 \\
        
        9 & T1083: File and Directory Discovery/DC~\cite{T1083} $\land$ T1033: System Owner/User Discovery/DC  $\implies$   T1082: System Information Discovery/DC  & 0.12 &    0.95 \\
        
        10 & T1057: Process Discovery/DC~\cite{T1057} $\land$ T1105: Ingress Tool Transfer/CC~\cite{T1105} $\land$ T1016: System Network Configuration Discovery/DC  $\implies$   T1082: System Information Discovery/DC & 0.12 &    0.95 \\
        \midrule
        \multicolumn{3}{l}{Total: 4670, minimum confidence: 0.16, mean confidence: 0.49, median confidence: 0.47} \\ \midrule
        \multicolumn{4}{p{15cm}}{IA: TA0001 - Initial access~\cite{TA0001}, EX: TA0002 - Execution~\cite{TA0002}, DC: TA0007 - Discovery~\cite{TA0007}, CC: TA0011 - Command and control~\cite{TA0002}, EF: TA0010 - Exfiltration~\cite{TA0010}, DE: TA0005 - Defense evasion~\cite{TA0005}} \\
    \bottomrule    
    \end{tabular}
    \label{tab:frequent-rules-comp}
\end{table*}

We identify six rules containing \textit{T1059: Command and Scripting Interpreter} technique. In all six rules, the technique is a consequent. The observation suggests the \textit{T1059: Command and Scripting Interpreter} technique may co-occur with a relatively high number of other adversarial techniques. Moreover, the other adversarial techniques implying the occurrence of \textit{T1059} indicate that \textit{T1059} has an overlapping attack vector with other adversarial techniques. For example, \textit{T1204: User execution} and \textit{T1059} both use command execution. These rules may indicate the order of adversarial actions, such as in Rule 7, where adversaries apply deobfuscation to obfuscated files before further usage. The rules can also suggest the usage of related techniques together: such as in Rule 2, where both techniques aid adversary in identifying system information. We also observe mutual interaction between two techniques, such as in Rule 1 and Rule 10, where phishing and user execution both implies the occurrence of each other. We also observe the transitive property~\cite{tanton2005encyclopedia} among the rules. Such as the Rule 1, 9, and 5, where, \textit{T1566} implies \textit{T1059} based on the fact that, \textit{T1566} implies \textit{T1204}, and \textit{T1204} implies \textit{T1059}. Seven of the ten rules have techniques from \textit{TA0002: Execution} tactic, which reflects that adversaries pick techniques that can eventually aid them in getting the opportunity to execute malicious commands in the victim environment. 

Table~\ref{tab:frequent-rules-comp} shows the Top 10 compound rules. For example, Rule 1 denotes that when an adversary uses both \textit{T1105: Ingress Tool Transfer} and \textit{T1566: Phishing}, then we can predict the adversary is also 99\% likely to use the \textit{T1204: User Execution}. In 12\% of the adversary entities, this rule materializes. While simple rules provide a one-to-one probability of the co-occurrence of two techniques, compound rules may provide a bigger context of how adversaries use techniques in conjunction. For example, Rule 3 shows that malicious tools used by adversaries are related to discovering system information. Rule 4 captures the context that an adversary has phished a user with an obfuscated malicious attachment which the user would later execute. Discussing all the thousands of simple and compound rules is out of the scope. However, cybersecurity researchers and practitioners can obtain and investigate all the rules from the replication package of this study. Organizations can use these rules as the starting point for the prediction of adversarial techniques. As Milajerdi et al. emphasized, false alarms of intrusion activity is one of the existing challenges in indicator-based threat detection systems~\cite{milajerdiHOLMESRealTimeAPT2019}. Co-occurrence rules can help in filtering malicious activity from benign ones.


\noindent \fbox{\begin{minipage}{24.5em}
\ding{43} \textbf{DS-4}: Organizations can predict potentially associated techniques from prior occurrence of techniques which can aid them detect and forecast intrusion activities with higher precision. 

\ding{43} \textbf{MS-3}: Organizations can enforce appropriate security controls and improve security practices upon the investigation of the underlying reasons for technique co-occurrences in their environment, such as dependency and similar attack vectors among techniques. 
\end{minipage}}


\begin{table*}[]
    \centering
    \caption{Adjacency matrix of the co-occurrence network. In the table, * denotes a directed edge exists from the technique located on row to the technique located on the column}
    \setlength\tabcolsep{2pt}
    
    \begin{tabular}{ll|l|lllll|ll|lllllllll|llllll|llll|lllll|l}
    \toprule
    
    {} & {} & IA & \multicolumn{5}{c}{EX} & \multicolumn{2}{c}{PS} & \multicolumn{9}{c}{DE} & \multicolumn{6}{c}{DC} & \multicolumn{4}{c}{CL} & \multicolumn{5}{c}{CC} & EF \\ \midrule
    
           \textbf{Tactic} & \textbf{Technique} & \begin{sideways} \textbf{T1566} \end{sideways}   & \begin{sideways} \textbf{T1047} \end{sideways}   & \begin{sideways} \textbf{T1053} \end{sideways}   & \begin{sideways} \textbf{T1059} \end{sideways}  & \begin{sideways} \textbf{T1106} \end{sideways}   & \begin{sideways} \textbf{T1204} \end{sideways}   & \begin{sideways} \textbf{T1543} \end{sideways}   & \begin{sideways} \textbf{T1547} \end{sideways}   & \begin{sideways} \textbf{T1027} \end{sideways}   & \begin{sideways} \textbf{T1036} \end{sideways}   & \begin{sideways} \textbf{T1055} \end{sideways}   & \begin{sideways} \textbf{T1070} \end{sideways}   & \begin{sideways} \textbf{T1112} \end{sideways}   & \begin{sideways} \textbf{T1140} \end{sideways}   & \begin{sideways} \textbf{T1218} \end{sideways}   & \begin{sideways} \textbf{T1562} \end{sideways}   & \begin{sideways} \textbf{T1564} \end{sideways}   & \begin{sideways} \textbf{T1016} \end{sideways}   & \begin{sideways} \textbf{T1033} \end{sideways}   & \begin{sideways} \textbf{T1057} \end{sideways}   & \begin{sideways} \textbf{T1082} \end{sideways}   & \begin{sideways} \textbf{T1083} \end{sideways}   & \begin{sideways} \textbf{T1518} \end{sideways}   & \begin{sideways} \textbf{T1005} \end{sideways}  & \begin{sideways} \textbf{T1056} \end{sideways}   & \begin{sideways} \textbf{T1113} \end{sideways}   & \begin{sideways} \textbf{T1560} \end{sideways}   & \begin{sideways} \textbf{T1071} \end{sideways}   & \begin{sideways} \textbf{T1090} \end{sideways}   & \begin{sideways} \textbf{T1105} \end{sideways}   & \begin{sideways} \textbf{T1132} \end{sideways}   & \begin{sideways} \textbf{T1573} \end{sideways}   & \begin{sideways} \textbf{T1041} \end{sideways}   \\
    \midrule
     
     IA & T1566: Phishing &         &         &         & *       &         & *       &         &         & *       &         &         &         &         &         &         &         &         &         &         &         &         &         &         &         &         &         &         &         &         & *       &         &         &         \\ \midrule
     
 EX & T1047: Windows Management Instrumentation~\cite{T1047} &         &         &         & *       &         &         &         &         &         &         &         &         &         &         &         &         &         &         &         &         &         &         &         &         &         &         &         &         &         &         &         &         &         \\
 
 EX & T1053: Scheduled Task/Job~\cite{T1053} &         &         &         & *       &         &         &         & *       & *       &         &         & *       &         &         &         &         &         &         &         &         & *       &         &         &         &         &         &         & *       &         & *       &         &         &         \\
 
 EX & T1059: Command and Scripting Interpreter~\cite{T1059} &         &         &         &         &         &         &         &         & *       &         &         &         &         &         &         &         &         &         &         &         & *       &         &         &         &         &         &         & *       &         & *       &         &         &         \\
 
 EX & T1106: Native API~\cite{T1106} &         &         &         & *       &         &         &         &         & *       &         &         & *       &         &         &         &         &         &         &         &         & *       & *       &         &         &         &         &         & *       &         & *       &         &         &         \\
 
 EX & T1204: User Execution~\cite{T1204} & *       &         &         & *       &         &         &         & *       & *       &         &         &         &         &         &         &         &         &         &         &         & *       &         &         &         &         &         &         & *       &         & *       &         &         &         \\ \midrule
 
 PS & T1543: Create or Modify System Process~\cite{T1543} &         &         &         & *       &         &         &         &         & *       &         &         & *       &         &         &         &         &         &         &         &         & *       &         &         &         &         &         &         & *       &         & *       &         &         &         \\
 
 PS & T1547: Boot or Logon Autostart Execution~\cite{T1547} &         &         &         & *       &         &         &         &         & *       &         &         & *       &         &         &         &         &         &         &         &         & *       &         &         &         &         &         &         & *       &         & *       &         &         &         \\ \midrule
 
 DE & T1027: Obfuscated Files or Information~\cite{T1027} &         &         &         & *       &         &         &         &         &         &         &         & *       &         &         &         &         &         &         &         &         & *       &         &         &         &         &         &         & *       &         & *       &         &         &         \\
 
 DE & T1036: Masquerading~\cite{T1036} &         &         &         & *       &         &         &         &         & *       &         &         &         &         &         &         &         &         &         &         &         & *       &         &         &         &         &         &         & *       &         & *       &         &         &         \\
 
 DE & T1055: Process Injection~\cite{T1055} &         &         &         & *       &         &         &         & *       & *       &         &         & *       &         &         &         &         &         &         &         & *       & *       & *       &         &         &         &         &         & *       &         & *       &         &         &         \\
 
 DE & T1070: Indicator Removal on Host~\cite{T1070} &         &         &         & *       &         &         &         &         & *       &         &         &         &         &         &         &         &         &         &         &         & *       & *       &         &         &         &         &         & *       &         & *       &         &         &         \\
 
 DE & T1112: Modify Registry~\cite{T1112} &         &         &         & *       &         &         &         &         & *       &         &         &         &         &         &         &         &         &         &         &         & *       &         &         &         &         &         &         &         &         & *       &         &         &         \\
 
 DE & T1140: Deobfuscate/Decode Files or Information~\cite{T1140} &         &         &         & *       &         &         &         &         & *       &         &         &         &         &         &         &         &         &         &         &         & *       & *       &         &         &         &         &         & *       &         & *       &         &         &         \\
 
 DE & T1218: Signed Binary Proxy Execution~\cite{T1218} &         &         &         & *       &         &         &         & *       & *       &         &         &         &         &         &         &         &         &         &         &         & *       &         &         &         &         &         &         & *       &         & *       &         &         &         \\
 
 DE & T1562: Impair Defenses~\cite{T1562} &         &         &         & *       &         &         &         &         &         &         &         &         &         &         &         &         &         &         &         &         &         &         &         &         &         &         &         &         &         &         &         &         &         \\
 
 DE & T1564: Hide Artifacts~\cite{T1564} &         &         &         & *       &         &         &         &         &         &         &         &         &         &         &         &         &         &         &         &         &         &         &         &         &         &         &         &         &         &         &         &         &         \\ \midrule
 
 DC: & T1016: System Network Configuration Discovery~\cite{T1016} &         &         &         & *       &         &         &         &         & *       &         &         & *       &         &         &         &         &         &         &         & *       & *       & *       &         &         &         &         &         & *       &         & *       &         &         &         \\
 
 DC: & T1033: System Owner/User Discovery~\cite{T1033} &         &         &         & *       &         &         &         &         & *       &         &         & *       &         &         &         &         &         & *       &         & *       & *       & *       &         &         &         &         &         & *       &         & *       &         &         &         \\
 
 DC: & T1057: Process Discovery~\cite{T1057} &         &         &         & *       &         &         &         &         & *       &         &         & *       &         &         &         &         &         & *       &         &         & *       & *       &         &         &         &         &         & *       &         & *       &         &         &         \\
 
 DC: & T1082: System Information Discovery~\cite{T1082} &         &         &         & *       &         &         &         &         & *       &         &         & *       &         &         &         &         &         & *       &         & *       &         & *       &         &         &         &         &         & *       &         & *       &         &         &         \\
 
 DC: & T1083: File and Directory Discovery~\cite{T1083} &         &         &         & *       &         &         &         &         & *       &         &         & *       &         &         &         &         &         &         &         & *       & *       &         &         &         &         &         &         & *       &         & *       &         &         &         \\
 
 DC: & T1518: Software Discovery~\cite{T1518} &         &         &         & *       &         &         &         &         & *       &         &         &         &         &         &         &         &         &         &         & *       & *       &         &         &         &         &         &         & *       &         & *       &         &         &         \\ \midrule
 
 CL: & T1005: Data from Local System~\cite{T1005} &         &         &         & *       &         &         &         &         & *       &         &         &         &         &         &         &         &         &         &         &         & *       & *       &         &         &         &         &         & *       &         & *       &         &         &         \\
 
 CL: & T1056: Input Capture~\cite{T1056} &         &         &         & *       &         &         &         & *       & *       &         &         & *       &         &         &         &         &         &         &         & *       & *       & *       &         &         &         & *       &         & *       &         & *       &         &         &         \\
 
 CL: & T1113: Screen Capture~\cite{T1113} &         &         &         & *       &         &         &         & *       & *       &         &         & *       &         &         &         &         &         &         &         & *       & *       & *       &         &         & *       &         &         & *       &         & *       &         &         &         \\
 
 CL: & T1560: Archive Collected Data~\cite{T1560} &         &         &         & *       &         &         &         &         & *       &         &         & *       &         &         &         &         &         &         &         &         & *       & *       &         &         &         &         &         & *       &         & *       &         &         &         \\ \midrule
 
 CC & T1071: Application Layer Protocol~\cite{T1071} &         &         &         & *       &         &         &         &         & *       &         &         &         &         &         &         &         &         &         &         &         & *       &         &         &         &         &         &         &         &         & *       &         &         &         \\
 
 CC & T1090: Proxy~\cite{T1090} &         &         &         & *       &         &         &         &         &         &         &         &         &         &         &         &         &         &         &         &         &         &         &         &         &         &         &         &         &         &         &         &         &         \\
 
 CC & T1105: Ingress Tool Transfer~\cite{T1105} &         &         &         & *       &         &         &         &         & *       &         &         &         &         &         &         &         &         &         &         &         & *       &         &         &         &         &         &         & *       &         &         &         &         &         \\
 
 CC & T1132: Data Encoding~\cite{T1132} &         &         &         & *       &         &         &         &         &         &         &         &         &         &         &         &         &         &         &         &         & *       &         &         &         &         &         &         & *       &         & *       &         &         &         \\
 
 CC & T1573: Encrypted Channel~\cite{T1573} &         &         &         & *       &         &         &         &         & *       &         &         &         &         &         &         &         &         &         &         &         & *       &         &         &         &         &         &         & *       &         & *       &         &         &         \\ \midrule
 
 EF & T1041: Exfiltration over C2 Channel~\cite{T1041} &         &         &         & *       &         &         &         &         & *       &         &         &         &         &         &         &         &         &         &         &         & *       & *       &         &         &         &         &         & *       &         & *       &         &         &         \\ \midrule
 
 \multicolumn{35}{p{16cm}}{IA: TA0001 - Initial access~\cite{TA0001}, EX: TA0002 - Execution~\cite{TA0002}, PS: TA0003 - Persistence~\cite{TA0003}, DE: TA0005 - Defense evasion~\cite{TA0005}, DC: TA0007 - Discovery~\cite{TA0007}, CL: TA0009 - Collection~\cite{TA0009}, CC: TA0011 - Command and control~\cite{TA0011}, EF: TA0010 - Exfiltration~\cite{TA0010}} \\
     
    \bottomrule
    \end{tabular}
    
    \label{tab:network-matrix}
\end{table*}

\subsection{Adversary Behavior}
\subsubsection{Co-occurrence network}
We obtain 376 simple co-occurrence rules, and the median confidence value of the rules is 0.51, as shown in Table~\ref{tab:frequent-rules-simple}. We build a co-occurrence network based on the simple rules having a confidence value of at least 0.51. In Table~\ref{tab:network-matrix}, we show the co-occurrence network in adjacency matrix format. The network has 33 nodes and  188 edges representing 188 simple co-occurrence rules among 33 techniques. We reflect on our findings from the network upon several following assumptions: (a) a directed edge from $technique_a$ to $technique_b$ implies that an adversary entity is likely to use $technique_b$ by at least 51\% whenever the entity uses $technique_a$; (b) the directed edge, however, cannot imply the order of the usage of technique - we cannot conclude $technique_b$ follows $technique_a$; (c) two techniques co-occurring together happens due to one or multiple reasons as discussed in~\cref{sec:rq2:1}. Investigating the reason(s) for each specific co-occurrence is out of the scope. However, a technique being a common consequent of multiple antecedent techniques may indicate that the antecedents have an interaction dependency on the consequent. A technique being a common antecedent of multiple consequent techniques, may indicate the antecedent technique determines the set of consequent techniques chosen by the adversary entities. We identify the adversary behaviors of techniques used from the observation of the assumptions in the paragraph. Below, we report our observation from the table, and upon the observation, we identify corresponding adversary behaviors (AV).
\begin{itemize}
    
    \item Techniques from \textit{DC: TA0007: Discovery} tactics co-occurs with other techniques from the same tactic most (n = 17). The observation indicates \textbf{AV-1}: adversaries deploy multiple means to discover system information
    
    \item \textit{DE: TA0005: Defense Evasion}, and \textit{CC: TA0011: Command and Control} is the pair of different tactics having the highest co-occurrence among their techniques (n=13). The observation indicates \textbf{AV-2}: adversaries utilize techniques of the two tactics together most: \textit{TA0005: Defense Evasion}, and \textit{TA0011: Command and Control}. The behavior may also indicate the adversaries aim to hide their footprint while communicating with adversary-controlled remote devices
    
    \item \textit{EX: TA0002: Execution} and \textit{CL: TA0009: Collection} have the highest number of consequent tactics (n = 6). The observation indicates \textbf{AV-3}: \textit{TA0002: Execution}, and \textit{TA0009: Collection} tactics mostly determine the follow-on tactics adversaries would perform through techniques
    
    \item All the seven tactics reported in the table can imply the occurrence of techniques from \textit{EX: TA0002: Execution}, \textit{DE: TA0005: Defense Evasion}, and \textit{CC: TA0011: Command and Control} tactics. The observation indicates \textbf{AV-4}: adversaries require to achieve the goal of \textit{TA0002: Execution}, \textit{TA0005: Defense Evasion}, \textit{TA0011: Command and Control} tactics to perform techniques from all other tactics reported in the table
    
    \item Techniques from \textit{DC: TA0007: Discovery} tactic have the highest number of consequent techniques (n = 46). The observation indicates \textbf{AV-5}: adversaries use techniques from \textit{TA0007: Discovery} to determine follow-on techniques
    
    \item Techniques from \textit{CC: TA0011: Command and Control} have the highest number of antecedent techniques (n = 54). The observation indicates \textbf{AV-6}:  the techniques from \textit{TA0011: Command and Control} tactic have the most interaction dependency with other techniques
    
    \item No techniques from the following six tactics appear in the table:   \textit{TA0004: Privilege escalation}, \textit{TA0006: Credential access}, \textit{TA0008: Lateral movement}, \textit{TA0040: Impact}, \textit{TA0043: Reconnaissance}, and \textit{TA0042: Resource development}. Techniques from these tactics did not appear as antecedents or consequents in any rule with 0.51 confidence value. In Table~\ref{tab:tactics-support}, we also observe that only three techniques exists from these six tactics and their support is also the lowest compared to techniques from other tactics. The observations indicate \textbf{AV-7}: \textit{TA0004: Privilege escalation}, \textit{TA0006: Credential access}, \textit{TA0008: Lateral movement}, \textit{TA0040: Impact}, \textit{TA0043: Reconnaissance}, and \textit{TA0042: Resource development} tactics are independent of other tactics. 
    
\end{itemize}

\begin{table*}[]
    \centering
    \caption{Centrality measures the techniques in the network}
    \label{tab:cent-values}
    \begin{tabular}{llrrrrr}
        \toprule
\textbf{Technique}                                     & \textbf{Tactic}              & \textbf{IDC} & \textbf{ODC} & \textbf{BC}  \\ \midrule

T1566: Phishing~\cite{T1566}                                        & TA0001: Initial Access~\cite{TA0001}               & 1            & 4            & 0                   \\ \midrule

T1047: Windows Management Instrumentation~\cite{T1047}              & TA0002: Execution~\cite{TA0002}                    & 0            & 1            & 0                               \\

T1053: Scheduled Task/Job~\cite{T1053}                              & TA0002: Execution~\cite{TA0002}                    & 0            & 7            & 0                               \\

\textbf{T1059: Command and Scripting Interpreter}~\cite{T1059}      & \textbf{TA0002: Execution}~\cite{TA0002}           & \textbf{32}  & 4            & \textbf{33.7}                     \\

T1106: Native API~\cite{T1106}                                      & TA0002: Execution~\cite{TA0002}                    & 0            & 7            & 0                              \\

T1204: User Execution~\cite{T1204}                         & TA0002: Execution~\cite{TA0002}           & 1            & 7            & 2.2                     \\ \midrule

T1543: Create or Modify System Process~\cite{T1543}                 & TA0003: Persistence~\cite{TA0003}                  & 0            & 6            & 0                               \\

T1547: Boot or Logon Autostart Execution~\cite{T1547}               & TA0003: Persistence~\cite{TA0003}                  & 6            & 6            & 0.67                              \\ \midrule

\textbf{T1027: Obfuscated Files or Information}~\cite{T1027}        & \textbf{TA0005: Defense Evasion}~\cite{TA0005}     & \textbf{27}  & 5            & \textbf{9.65}                    \\

T1036: Masquerading~\cite{T1036}                                    & TA0005: Defense Evasion~\cite{TA0005}              & 0            & 5            & 0                              \\

T1055: Process Injection~\cite{T1055}                      & TA0005: Defense Evasion~\cite{TA0005}     & 0            & \textbf{9}   & 0              \\

T1070: Indicator Removal on Host~\cite{T1070}                       & TA0005: Defense Evasion~\cite{TA0005}              & 14           & 6            & 2.2                               \\

T1112: Modify Registry~\cite{T1112}                                 & TA0005: Defense Evasion~\cite{TA0005}              & 0            & 4            & 0                             \\

T1140: Deobfuscate/Decode Files or Information~\cite{T1140}         & TA0005: Defense Evasion~\cite{TA0005}              & 0            & 6            & 0                                \\

T1218: Signed Binary Proxy Execution~\cite{T1218}                   & TA0005: Defense Evasion~\cite{TA0005}              & 0            & 6            & 0                              \\

T1562: Impair Defenses~\cite{T1562}                                 & TA0005: Defense Evasion~\cite{TA0005}              & 0            & 1            & 0                                \\

T1564: Hide Artifacts~\cite{T1564}                                  & TA0005: Defense Evasion~\cite{TA0005}              & 0            & 1            & 0                            \\ \midrule

\textbf{T1016: System Network Configuration Discovery}~\cite{T1016} & \textbf{TA0007: Discovery}~\cite{TA0007}           & 3            & \textbf{8}   & 0               \\

\textbf{T1033: System Owner/User Discovery}~\cite{T1033}            & \textbf{TA0007: Discovery}~\cite{TA0007}           & 0            & \textbf{9}   & 0              \\

\textbf{T1057: Process Discovery}~\cite{T1057}                      & \textbf{TA0007: Discovery}~\cite{TA0007}           & 8            & \textbf{8}   & \textbf{3.33}   \\

\textbf{T1082: System Information Discovery}~\cite{T1082}           & \textbf{TA0007: Discovery}~\cite{TA0007}           & \textbf{27}  & \textbf{8}   & \textbf{72.3}   \\

\textbf{T1083: File and Directory Discovery}~\cite{T1083}           & \textbf{TA0007: Discovery}~\cite{TA0007}           & 13           & 7            & \textbf{4}                        \\

T1518: Software Discovery~\cite{T1518}                              & TA0007: Discovery~\cite{TA0007}                    & 0            & 6            & 0                              \\ \midrule

T1005: Data from Local System~\cite{T1105}                          & TA0009: Collection~\cite{TA0009}                   & 0            & 6            & 0                              \\

\textbf{T1056: Input Capture}~\cite{T1056}                          & \textbf{TA0009: Collection}~\cite{TA0009}          & 1            & \textbf{10}  & 0              \\

\textbf{T1113: Screen Capture}~\cite{T1113}                         & \textbf{TA0009: Collection}~\cite{TA0009}          & 1            & \textbf{10}  & 0               \\

T1560: Archive Collected Data~\cite{T1560}                          & TA0009: Collection~\cite{TA0009}                   & 0            & 7            & 0                             \\ \midrule

T1041: Exfiltration Over C2 Channel~\cite{T1041}                    & TA0010: Exfiltration~\cite{TA0010}                 & 0            & 6            & 0                              \\ \midrule

\textbf{T1071: Application Layer Protocol}~\cite{T1071}             & \textbf{TA0011: Command and Control}~\cite{TA0011} & \textbf{26}  & 4            & 0.25                             \\

T1090: Proxy~\cite{T1090}                                           & TA0011: Command and Control~\cite{TA0011}          & 0            & 1            & 0                           \\

\textbf{T1105: Ingress Tool Transfer}~\cite{T1105}                  & \textbf{TA0011: Command and Control}~\cite{TA0011} & \textbf{28}  & 4            & 1.7                             \\

T1132: Data Encoding~\cite{T1132}                                   & TA0011: Command and Control~\cite{TA0011}          & 0            & 4            & 0                    \\

T1573: Encrypted Channel~\cite{T1573}                               & TA0011: Command and Control~\cite{TA0011}          & 0            & 5            & 0                      \\ \midrule

\multicolumn{7}{p{14cm}}{IDC: in-degree centrality, ODC: out-degree centrality, BC: betweenness centrality} \\
        \bottomrule
        \end{tabular}
\end{table*}

\subsubsection{Centrality Measures}
We compute the three centrality measures (Table~\ref{tab:cent-def}) on the co-occurrence network, and we report the scores for each of the 33 techniques in Table~\ref{tab:cent-values}. For each of the centrality measures, we bold the top five techniques and any techniques having equal value to the fifth topmost technique. We discuss our observations from the table in the following sections.

\subsubsection{In-degree (IDC) and out-degree (ODC) centrality} 
IDC of a technique $te$ indicates how many other techniques can imply the occurrence of $te$. ODC of a technique $te$ indicates the occurrence of how many techniques $te$ can imply. We observe the following from Table~\ref{tab:cent-values}, and we identify adversary behaviors upon the observation.

\textit{T1059: Command and Scripting Interpreter} has the highest in-degree centrality, and the technique has incoming edges from all the other 32 techniques. The four next topmost technique in in-degree centrality are: \textit{T1105: Ingress Tool Transfer} (n=28), \textit{T1027: Obfuscated File/Information} (n=27), \textit{T1082: System Information Discovery} (n=27), and \textit{T1071: Application Layer Protocol} (n=26). The top five techniques have incoming edges from at least 78\% of the techniques in the network. However, 19 techniques do not have any incoming edges at all. The observation indicates a relatively high number of techniques depends on the top five techniques in in-degree centrality. 

We observe that these techniques are rather an abuse of legitimate system functionality. All operating systems (OS) and platforms ship with built-in command execution capabilities, which adversaries abuse through \textit{T1059: Command and Scripting Interpreter}. OS and platforms also can communicate with remote systems through various application layer protocols such as FTP, HTTPS, and SSH. Adversaries abuse the capability to import malicious files and tools through \textit{T1105: Ingress tool transfer} using \textit{T1071: Application layer protocol}. All OS and platforms also facilitate encryption or encoding, which aids adversaries in obfuscating malicious files or their traces through \textit{T1027: Obfuscated files or information}. Adversaries also identify critical software and hardware information along with potential vulnerabilities and patches of the victim environment through \textit{T1082: System information discovery}. Overall, the observation indicates \textbf{AV-8}: adversaries primarily abuse the following legitimate system functionality: command line interface, application layer protocol, file transfer, encoding/encryption, and read system properties. A relatively high number of other malicious techniques depend on abusing these five functionalities. 

We observe from Table~\ref{tab:cent-values} that \textit{T1056: Input Capture} and \textit{T1113: Screen Capture} have the highest out-degree centrality (n=10). The five next topmost technique in out-degree centrality are: \textit{T1033: System Owner/User Discovery} (n=9), \textit{T1055: Process Injection} (n=9), \textit{T1016: System Network Configuration Discovery} (n=8), \textit{T1057: Process Discovery} (n=8), and \textit{T1082: System Information Discovery} (n=8). We also observe, through six out of the top seven techniques in the out-degree centrality, adversaries attempt to understand aspects of the victim system such as architecture, network, file system, or collect the information of interest. The observation suggests \textbf{AV-9}: an adversary's follow-on techniques depend on the obtained information of victim systems. We also observe that while many techniques have no incoming edges, all have at least one outgoing edge. The observation indicates \textbf{AV-10}: while the majority of the techniques do not act as a precursor for other techniques, all the techniques can imply adversaries' follow-on choice of techniques. Finally, we observe \textit{T1082: System Information Discovery} technique is the only technique among the top five in both in-degree and out-degree centrality. The observation indicates \textbf{AV-11}: \textit{T1082: System Information Discovery} acts as a common antecedent and consequent both. The technique aids adversaries by acting as a precursor for implementing other techniques. Moreover, the technique also shapes adversaries' follow-on behavior. 

The top techniques in in-degree and out-degree centrality reflect the local aspect of malicious actions. For example, the incoming and outgoing edges from \textit{T1059} reflect how an adversary can use the command line in conjunction with other adversarial techniques. Thus the in-degree and out-degree centrality can aid practitioners in detection and mitigation by the following:

\noindent \fbox{\begin{minipage}{24.5em}
\ding{43} \textbf{DS-5}: Organizations can build detection rules around the following techniques: \textit{T1016, T1027, T1033, T1056, T1057, T1059, T1071, T1082, T1105, T1113}, aiding them in detecting many other correlated techniques. 

\ding{43} \textbf{MS-4}: The following techniques: \textit{T1016, T1027, T1033, T1056, T1057, T1059, T1071, T1082, T1105, T1113} - are associated with many techniques. Mitigation of these can eventually aid in mitigating the associated techniques. 

\ding{43} \textbf{MS-5} Organizations should investigate why specific techniques are highly associated with others and whether such associations are correlated with the limitations of security solutions and practices. For example, suppose an organization finds \textit{T1059:  Command and Scripting Interpreter} is associated with many adversarial techniques. The organization may investigate whether the system configuration allows the non-restricted command execution in the environment. 
\end{minipage}}

\subsubsection{Betweenness Centrality (BC)} 
BC of a technique $te$ indicates to what extent $te$ serves as a bridge between one group of techniques to another. High betweenness captures the context that the occurrence of $te$ would lead to the occurrence of $te$'s antecedent and consequent technique with greater probability than the case if $te$ does not occur. We observe the following from Table~\ref{tab:cent-values}: (a) only 10 out of 33 have a non-zero value in this centrality. Among the 10, only \textit{T1082: System Information Discovery} (n=72.3), and \textit{T1059: Command and Scripting Interpreter} (n=33.7) have a relatively high score. 

The BC score of these two techniques suggests that \textit{T1082} and \textit{T1059} technique lies on 72\% and 33.7\% respectively of all the shortest chained rules among all pairs of techniques in the network. Thus, betweenness centrality may reflect how adversaries can use techniques together in a sequence. For example, \textit{T1082} lies on the shortest chained rule between these two techniques: \textit{T1105: Ingress Tool Transfer}, and \textit{T1070: Indicator Removal on Hosts}. The example indicates that the adversary transfers malicious contents into the victim system from a remote and hides its traces. However, the adversary uses \textit{T1082} to figure out the victim environment to determine how to hide the footprints of the malicious content. The observation indicates \textbf{AV-12}: The \textit{T1082: System Information Discovery}, and \textit{T1059: Command and Scripting Interpreter} techniques aid an adversary in understanding and broadening the attack surface. These two techniques, thus, may act as common steps in different possible sequences of techniques.  

\noindent \fbox{\begin{minipage}{24.5em}
\ding{43} \textbf{DS-6}: Practitioners can build detection rules around command line execution and reading system properties to capture potential sequences of adversarial techniques. 

\ding{43} \textbf{MS-6} By mitigating only the two techniques: \textit{T1082} and \textit{T1059}, organizations can disrupt adversaries' sequence of malicious actions.\end{minipage}}

\section{Discussion and future research directions}
\label{sec:discussion}
\textbf{What adversarial techniques require the most attention?} Although MITRE ATT\&CK enlists 188 techniques, we observe only 44 of those are prevalent among at least 10\% of the adversaries. \textit{T1059: Command and control}, \textit{T1105: Ingress tool transfer}, \textit{T1027: Obfuscated files or information}, \textit{T10721: Application layer protocol}, and \textit{T1082: System Information Discovery} techniques are the predominant techniques reported in 50\% of the cybercrime groups and malware. Adversaries pair these five techniques with a large number of ATT\&CK techniques. Hence, these techniques work like hubs for other techniques -- as adversaries can utilize many other techniques if they can apply these five. Although these five techniques are abuses of system functionality, we observe three means of abuse. Through \textit{T1105: Ingress tool transfer}, \textit{T1027: Obfuscated files or information} techniques, adversaries performs a write/execute operation to gain tactical objectives. However, adversaries exploit the built-in features of an OS through \textit{T1059: Command and control} and \textit{T10721: Application layer protocol} techniques. On the other hand, through \textit{T1082: System information discovery} technique, adversaries read the system/network information. As per ATT\&CK, \textit{TA0007: Discovery}-related techniques shown in Table~\ref{tab:top-ten-technique-support} do not have any mitigation at all. Two common mitigation techniques exist for the top four in the table: using anti-malware systems and network intrusion detection. Hence, enforcing these two mitigation techniques should be prioritized by defenders.

\textbf{Both defense-in-depth and defense-in-breadth are required.} Modern information technology infrastructures use a myriad of devices and thus create a heterogeneous environment consisting of network devices, servers, personal computers, mobile devices, and applications. Thus, enforcing security controls on only a subset of devices or applications may leave opportunities for adversaries to exploit elsewhere where security enforcement is insufficient. For example, an adversary can use \textit{T1059: Command and scripting interpreter} on personal computers, routers, and even mobile devices. Thus, an organization may enforce the strongest security measures on execution privileges on personal computers -- which is an example of defense in depth. However, an adversary may trick a user into executing a command in an unprotected mobile device -- which is an example of an adversary taking advantage due to the lack of defense in breadth. Thus, organizations require a synchronized effort to detect and mitigate techniques across all endpoints so that adversaries cannot circumvent the security enforcement of one facet of an environment by exploiting another unprotected one.    

\textbf{Machine learning:} We identify a set of co-occurrence rules among the techniques which can pave the path for building probabilistic models for predicting and detecting techniques, such as naive Bayesian or hidden Markov models. Intrusion detection systems generally suffer from false alert issues~\cite{milajerdiHOLMESRealTimeAPT2019}. These rules can aid these systems with greater precision in generating alerts. 

\textbf{Detection and mitigation strategies should be adaptive to accommodate the change in the threat landscape.} Our study identifies co-occurrence rules of techniques documented by MITRE ATT\&CK. However, the threat landscape changes and techniques used by adversaries also depend on the adversaries' expertise and the weakness of the victim environment. Hence, organizations can run co-occurrence analysis on their environment and adapt detection and mitigation strategies that suit organizations' workflow, system architecture, and security enforcement. Organizations can also conduct longitudinal analysis to discover the correlation between adversarial techniques and the limitations of security solutions.

\textbf{Cyberattack data should be made available for open research.} One major challenge in developing a co-coordinated effort to thwart adversaries is that companies are reluctant to openly share and distribute the knowledge on how system compromise happened and what security enforcement failed~\cite{cyberNTSB}. However, the lack of fact-finding and sharing eventually deters organizations from capturing a better and actionable understanding of protecting themselves from attacks. Based on our findings, we emphasize that organizations should share pre- and post-compromise facts so that independent researchers and cybersecurity vendors can collaborate to build an actionable knowledge base of cyberattack incidents, indicators, behaviors, and pathways for creating a secure environment. 

\textbf{Deeper investigation is required for technique co-occurrences.} Our study is built upon the primary assumption that frequently co-occurring techniques may correlate. However, deeper analysis is required to uncover the underlying reasons for the observed correlation, such as dependency, similar attack vectors, or flow along a sequence among the execution of techniques. Such investigation could be crucial for enforcing and enhancing security measures to prevent adversaries from using the techniques.


\section{Limitations}
\label{sec:threats}

We discuss several limitations of the study in this section. The dataset we use reflects the adversarial techniques documented by MITRE ATT\&CK. The dataset captures only the techniques used by 115 cybercrime groups and 484 malware. Moreover, the dataset contains only the techniques identified by security experts and then reported after cyberattacks. Consequently, adversaries may have used other techniques that were not detected and remain unreported. Thus, the dataset only reflects a subset of the techniques adversaries used. MITRE ATT\&CK mapped the techniques from publicly-reported documents using an automated manner~\cite{mitre-tram} - which may have introduced mapping bias in the dataset. Overall, the study's findings reflect the context of technique co-occurrence data documented by the ATT\&CK framework. Nonetheless, the study approximates how cybercrime groups and malware are leveraging techniques on various types of organizations across the globe. Practitioners and defenders can use the methodology and findings of the paper as a starting point to investigate adversarial techniques in their environment, derive mitigation strategies, and improve security practices. Hence, the approach presented in the study requires further data collection and ground truth construction from numerous types of organizations across various parts of the globe. The coordinated effort among organizations and cybersecurity vendors can capture generalized information on how techniques interact with one another and their probability distribution to co-occur. 

\section{Related Work}
\label{sec:related}
In this section, we discuss our related work. \textbf{Prediction}: In~\cite{sentunaNovelEnhancedNaive2021}, the authors proposed a machine learning model to predict a set of probable TTPs based upon the prior probabilities of detected TTPs in an environment. They trained the model with a Bayesian probability network~\cite{russell2010artificial} of TTPs obtained from reports describing cyberattacks. They evaluated their model on five datasets obtained from NSL-KDD~\cite{nslkdd}, CICIDS2017~\cite{cicids}, and MITRE ATT\&CK, and they obtained a prediction accuracy of 93-97\%. \textbf{Association among TTPs}: In~\cite{al-shaerLearningAssociationsMITRE2020}, the authors investigated whether the TTPs demonstrate correlation in cyberattack incidents. They obtained a dataset of 66 APT and 204 software attacks from MITRE ATT\&CK. They applied hierarchical clustering to group similar TTPs and identified 37 and 61 clusters for APT and software attacks. \textbf{Association among indicators and TTPs} In~\cite{abuAttributionCyberattackUsing2020}, the authors investigated the relationship between Indicators of Compromise (IoC)~\cite{iocdef} and malware from network traffic data. They applied an association rule mining algorithm and identified association rules among malware and corresponding traces, such as the Mirai~\cite{mirai} malware's strong association with the following IP: 212.61.180.100. 

\textbf{Threat model} In~\cite{choiProbabilisticAttackSequence2021}, the authors constructed a dataset of attack sequences on industrial control systems. The authors trained a hidden Markov model reflecting probabilistic transition among MITRE ATT\&CK tactics and techniques. In~\cite{karunaAutomatingCyberThreat2021}, the authors proposed an automatic threat hunting model based upon genetic programming. Using their proposed domain-specific language, the authors transformed cyberattack descriptions of MITRE ATT\&CK into TTPs. Then they applied genetic programming to obtain variations of a set of TTPs from known threat descriptions. They stored these variations in a database to facilitate querying and threat hunting for unknown threats. In~\cite{beradyTTPIoCAdvanced2021}, the authors proposed formalized advanced persistent attacks from the perspective of both attackers and defenders. They construct two graphs for attackers and defenders reflecting adversarial and defensive actions performed on a common set of objects in an environment. The authors then demonstrated how the model could be used to (a) unify the perception of both attackers and defenders; (b) identify traces of attacks on the objects; and (c) improve the threat hunting by measuring the success of defensive actions. The authors evaluated the model on simulated attacks mimicking cyberattacks launched by APT29 actors. 

\textbf{Threat actor attribution} In~\cite{kimAutomaticallyAttributingMobile2021}, the authors attributed threat actors from the ATT\&CK TTPs of mobile malware. The authors used a dataset of 120 mobile malware and 12 threat actors. For each malware, the authors computed the cosine similarity score of the malware based on the observed TTPs and IoCs. The similarity score is used as a distance function for cluster formation, where each cluster represents a responsible actor group. Their proposed attribution model shows 82\% and 90\% precision and recall scores. In~\cite{weiDeepHunterGraphNeural2021a}, the authors used graph embedding to represent known APT attacks. The authors then attributed malicious behaviors to known APT attacks through graph matching. They evaluated their proposed model on five real-world APT attacks where the model showed a 0.95 AUC score. In~\cite{wuGroupTracerAutomaticAttacker2020}, the authors proposed GroupTracer to identify the responsible actors of cyberattacks launched on Internet of Things (IoT) devices. GroupTracer constructs TTPs profile of attacks performed on IoT devices by responsible groups. Based upon the profile, GroupTracer performs hierarchical clustering on the TTPs profile of unseen attacks on IoT and finds potentially-accountable groups. 

\textbf{Association among indicators and TTPs} In~\cite{huangOpenSourceIntelligence2021}, the authors utilized MITRE ATT\&CK to identify TTPs and associated resources from the Windows malware execution trace. They proposed a machine learning model named MAMBA to automatically identify the TTPs along with corresponding API and resources. Their model showed over 90\% score in precision and recall. In~\cite{fairbanksIdentifyingATTCK2021}, the authors investigated the execution flow of Android malware from the perspective of MITRE ATT\&CK. They proposed a graph-based neural network-based learning model to identify specific TTPs from malware control flow graphs. They evaluated the model on 3,250 malware APKs, and the model achieved a 93\% F1 score. In~\cite{milajerdiHOLMESRealTimeAPT2019}, the authors mapped the intrusion activities to cyber kill-chain~\cite{ckc} stages and identified correlation among log messages involving different stages and resources of attacks. Based upon the mapping and correlation, the authors constructed attack graphs of various stages of APT attacks, and they evaluated their proposed model on nine real-world APT attack scenarios. In~\cite{takeyRealTimeEarly2021}, the authors proposed a model for detecting multi-stage attacks in an attack's early phases. Their proposed model can predict a malware's behavior and identify corresponding MITRE ATT\&CK techniques. In~\cite{hassanTacticalProvenanceAnalysis2020}, the authors proposed a tactical provenance graph based on MITRE ATT\&CK that represents causal dependencies among threat alerts. The authors showed that their proposed model could help practitioners reduce false alerts by their proposed alert scoring mechanism and preserve the semantics of threat alerts/log data using limited memory.  

Overall, in the literature, researchers utilized MITRE ATT\&CK for threat modeling, threat attribution, profiling malware actions, and studying the association among TTPs with indicators and intrusion alerts. In this work, we utilize ATT\&CK to understand how adversaries leverage multiple ATT\&CK techniques and how the understanding can drive proactive detection and mitigation strategies. 

\section{Dataset and source code}
\label{dataset}
The dataset and analysis scripts are available at: https://github.com/brokenquark/ttps-co-occurrence.

\section{Conclusion}
\label{sec:conclusion}
This study aims to understand how attackers use adversarial techniques in conjunction to enable organizations to formulate detection and mitigation strategies. To this end, we collect the publicly-reported MITRE ATT\&CK techniques from 599 cybercrime groups and malware. We obtain co-occurrence rules among the techniques through association rule mining and identify adversary behaviors using network analysis. We make our dataset and analysis scripts available for researchers to rerun the analysis on future versions of ATT\&CK and different datasets. Organizations can use our approach as a starting point to formulate proactive defensive strategies for protecting their environments. We also advocate researchers and practitioners make synchronized efforts to collect more data to draw a comprehensive picture of how cyberattack happens and how we can make organizations more secure.

\section*{Acknowledgment}
The authors would like to thank [redacted to protect anonymity.]

\bibliographystyle{plain}
\bibliography{main}

\end{document}